\documentclass[12pt,preprint]{aastex}

\shorttitle{EVOLUTION OF REST-FRAME COLOR OF GALAXIES}
\shortauthors{KAJISAWA \& YAMADA}

\begin{document}


\title{Evolution of the Dependence of Rest-frame Color and Morphology
Distribution on Stellar Mass for Galaxies in the Hubble Deep Field North}


\author{M. Kajisawa 
and T. Yamada
}
\affil{National Astronomical Observatory of Japan,\\ 
2-21-1, Osawa, Mitaka, Tokyo 181-8588, Japan}
\email{kajisawa@optik.mtk.nao.ac.jp}







\begin{abstract}
Using the Subaru very deep $K'$-band imaging and
HST WFPC2/NICMOS archival data of the Hubble Deep Field North,
we investigate the evolution of the stellar mass, color, morphology
 of galaxies to $z\sim3$. We mainly examine the rest-frame
$U-V$ color distribution of galaxies as a function of 
stellar mass. At $0.3\lesssim z\lesssim2$, galaxies
seem to be divided into the 
two populations at 
around the stellar mass of $\sim 5\times 10^{9}$ M$_{\odot}$.
The low-mass galaxies have relatively bluer rest $U-V$ color and
 their color does not show clear correlation with stellar
mass over the range of 10$^{8}$--5$\times$10$^{9}$M$_{\odot}$.
On the other hand, at higher mass, 
the more massive galaxies
 tend to have the redder $U-V$ color.
The average $U-V$ color of the low-mass galaxies becomes
bluer gradually with redshift, from $U-V \sim 0.2$ at $z \sim 0.5$ to
$U-V \sim -0.2$ at $z\sim 2$. On the contrary,
the correlation between the
stellar mass and rest $U-V$ color of the high-mass
population does not seem to change significantly 
between $z\sim0.3$ and $z\sim2$.
The morphological distribution shows that at $z\lesssim1$,
the low-mass population is dominated by disk galaxies,
while the fraction of early-type galaxies is larger 
 in the high-mass population.
At $1<z<2$, although the fraction of irregular galaxies increases,
the similar trend is observed.
At $z>2$, it is seen that more massive
 galaxies tend to have redder $U-V$ color over the range of
$10^{9}$-$10^{10}$M$_{\odot}$, although
we can only sample the galaxies with stellar mass larger than
$\sim 1\times10^{9}$M$_{\odot}$.
These results suggests that the star formation history of galaxies
depends on their stellar mass very much. The low-mass
population is likely to have relatively long star formation 
timescale, and their formation redshifts do not seem to be much higher than
$z\sim2$ under the assumption of constant star formation rate. 
At the stellar mass larger than
$\sim 5\times 10^{9}$M$_{\odot}$, 
there must be some mechanisms which suppress 
the star formation in galaxies at $0<z<2$.
\end{abstract}





\section{Introduction}

For the study of the galaxy formation in the framework of the
structure formation of the universe, it is important to investigate
the various properties of galaxies as a function of their mass.
Unlike color or morphology, since the dynamical or stellar mass is
relatively unaffected by recent star formation activities or the
morphological transformation processes such as interactions or
mergers, which increase mass at most factor of two,  
it can be conveniently used as independent variable
for the investigation of the star formation history 
or the evolution of the morphology distribution of galaxies.
While the dynamical mass is more directly connected to the prediction
of the structure formation theory, the stellar mass can be more easily
measured using the NIR luminosity with optical-NIR colors or
spectral stellar population indices (e.g, \citealp{gia98,bri00,kau03}).
Since the stellar mass of galaxies reflects the past history of
star formation, its evolution is closely related to 
the time when the majority of
stars in those galaxies built up.
Thus the investigations of the distribution of various properties of
galaxies as a function of stellar mass at various epochs
help our understanding of
the processes of galaxy formation and evolution.

For local universe, \citet{kau03} investigated the stellar
mass dependence of the star formation history, sizes, and structural
parameter of galaxies with the Sloan Digital Sky Survey data,
and found that galaxies divide into two
distinct families at a stellar mass of 3$\times 10^{10}$M$_{\odot}$.
These families are found to have  different star formation activities, 
morphologies, surface mass densities, and so on. 

At intermediate redshifts, \citet{gia98} estimated the
stellar mass of galaxies with $R'<25$ in the field of the $z=4.7$ quasar
BR 1202-0725 using the $BVrIK$ imaging data.
They found that the distribution of the
stellar mass of galaxies at $z=0.4$-0.8 has a median
value of $\sim5\times10^{8}$M$_{\odot}$, but the blue
galaxies with $B'-V'\leq1.4$ show a narrower and peaky stellar mass
distribution at $\sim10^{8}$M$_{\odot}$, and the all
galaxies with M$_{stellar}>2\times10^{9}$M$_{\odot}$ have
$B'-V'>1.4$.
\citet{bri00} derived the stellar mass of $I<22$
galaxies with spectroscopic redshifts in the CFRS and LDSS surveys
using the optical--NIR photometries, and investigated their relation
with morphology or star formation rate. They found that in the range
of $10.5 <\log{{\rm M}_{stellar}}<11.6$, the comoving stellar mass density in
irregular galaxies increases rapidly from $z\sim0.2$ to $z\sim1.0$,
and this is complemented by the modest decrease of the mass density in
spheroidal galaxies. They also pointed out that at $0.2<z<1.0$, the
more massive galaxies tend to have lower specific star formation rate
R$=$SFR/M$_{stellar}$. 

For galaxies at $z>2$, \citet{saw98} investigated
the broad-band optical--NIR SED of Lyman break
galaxies with spectroscopic redshifts in the Hubble Deep Field North
(HDF-N), and found that their stellar masses are about 1/15 of a
present-day L$^{*}$ galaxy and that these galaxies are dominated by
very young ($\lesssim0.2$Gyr) stellar population. 
\citet{pap01} also investigated the spectral
stellar population properties of those 
galaxies with spectroscopic redshifts
in the HDF-N, using the optical--NIR HST imaging data. They found that
the Lyman break galaxies with ``L$^{*}_{\rm LBG}$'' UV luminosity have the
stellar mass of roughly 1/10 of a present-day L$^{*}$ galaxy
($\sim10^{10}$M$_{\odot}$), and they are well fitted by the model
with population ages that range from 30 Myr to 1 Gyr. 
\citet{fon03} estimated the stellar mass of $z>2$
galaxies with $K<25$ using the HDF-S optical--NIR imaging data.
In addition to the star forming galaxies, they found that there are a
few red galaxies at $z\gtrsim2$, and that 
the fraction of older objects seems to increase at large stellar mass, 
although
the biases against old/passive objects exist near the detection limit.
\citet{sha01} investigated the stellar mass of relatively
bright $R<25.5$ Lyman break galaxies at $z\sim3$, using $UGRIK$-bands
photometries.
They found that their $R-K$ color distribution is wide-spread, and
their UV luminosity (both dust-corrected and uncorrected) is
uncorrelated with the estimated stellar mass.

On the other hand, several studies indicate that the morphological
sequence of galaxies seen in the present universe seems to be formed
between $z\sim1$ and $z\sim3$ (e.g., \citealp{dic00a,kaj01}).
The possible rapid evolution of cosmological stellar mass density 
at $1<z<3$ is also reported (\citealp{dic03,fon03}).
Therefore it is interesting to investigate the evolution of
the mass dependence of these properties of galaxies back to such an 
important epoch for galaxy formation.

  In this paper, in order to understand the formation and evolution of 
field galaxies, we investigate the rest-frame color and morphological 
distribution of galaxies in the Hubble Deep Field North 
as a function of stellar mass back to $z\sim3$,
using the archival HST WFPC2/NICMOS data and the very deep
Subaru/CISCO $K'$-band imaging. 
Deep $U$-to-$K$ band images allow us to investigate 
these properties of galaxies 
at the rest-frame $V$-band toward $z\sim3$, and 
estimate the rest $U-V$ color without extrapolation. 
The $U-V$ color is  relatively sensitive to the average age
of their stars because it covers the wavelength region
where the largest features in galaxy continuum
spectra at near-UV to optical wavelengths, namely, a 4000 \AA\
break and/or the Balmer discontinuity exist.
The wide wavelength coverage also help us to constrain the stellar
mass of galaxies with relatively high accuracy.
 In section 2, we describe the data reduction and the sample
selection, and the way to estimate the stellar mass, rest-frame color,
and morphology of our sample.
We investigate the evolution of the rest-frame $U-V$ color and
morphological distribution of galaxies as a function of 
  stellar mass in section 3. A discussion 
of these results is presented in section 4. In section 5, we summarize
the results obtained from our analysis.
We use the standard (Vega) magnitude system throughout the paper.
HST filter bands refer to
F300W, F450W, F606W, F814W, F110W, F160W bands as $U_{300}$,
$B_{450}$, $V_{606}$, $I_{814}$, $J_{110}$, $H_{160}$,
respectively.

\section{Data Reduction and Sample Selection}
\subsection{CISCO $K'$-band data}

 The HDF-N field was observed at the $K'$-band with the Subaru
telescope equipped with the Cooled Infrared Spectrograph and
Camera for OHS (CISCO, \citealp{mot02}) on 2001 April 3 and 4.
The detector used was a 1024 $\times$ 1024 HgCdTe array with a pixel
scale of 0.111 arcsec for the Optical Nasmyth secondary, which
provides a field of view of $114'' \times 
114''$. A number of 
dis-registered images with short exposures (20 s for each frame) were
taken in a circular dither pattern with a $10$ arcsec diameter. A series of
twelve frames were taken at each place and the telescope was then
moved to the next position. The position of each cycle (12 $\times$ 10
frames) was randomly dithered with about $10$ arcsec offset.
The total exposure time was 36740 sec, about 10 hours.
The weather condition was stable during the observations, and
seeing was between 0.3 and 0.7 arcsec, typically $\sim$ 0.5 arcsec.

\begin{figure}
\epsscale{0.65}
\plotone{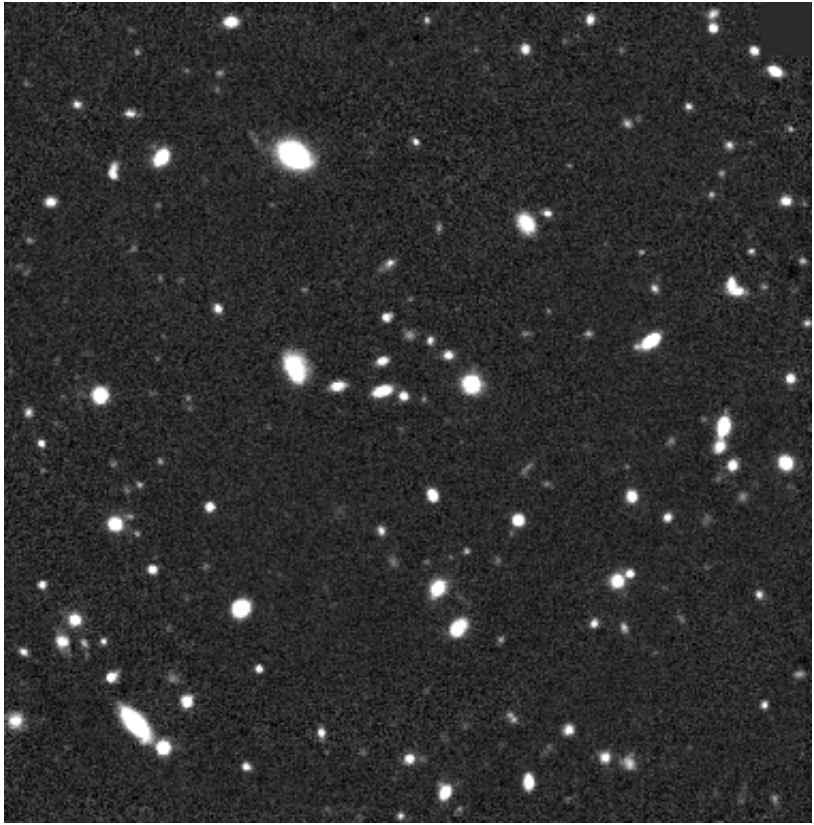}
\caption{CISCO $K'$-band image of the Hubble Deep Field North.
\label{HDFcisco}}
\end{figure}

\begin{figure}
\epsscale{0.45}
\plotone{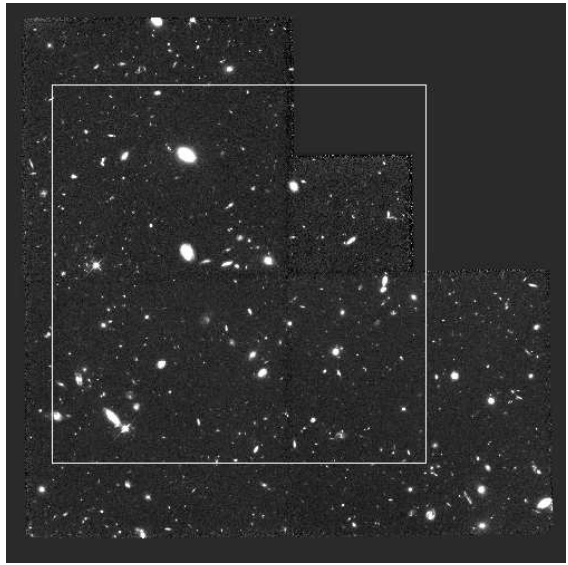}
\caption{Field of view of
CISCO $K'$-band image relative to HST WFPC2 image.
\label{FOVcisco}}
\end{figure}

  The data were reduced using the IRAF software package. There was a
variation in the dark level of CISCO, which depends on the sky
background level. At first, we thus performed flat-fielding using
the superflat frame for CISCO $K'$-band data \citep{mot02}, and measured the
background level of each frame. Then, ``sky background $+$ dark''
frame for each image was produced from the sample of frames with
nearly same background level, and the sky $+$ dark subtraction was performed.
Further we subtracted the residual background by fitting the frames
with 5th-order surface function. Then source detection was done for
each frame and the offsets between frames were measured.
These frames were co-registered and combined. We then performed
tentative source detection for this combined frame to make the mask
image. Using this mask image, we masked objects in each raw frame,
and remade ``sky background $+$ dark'' frame, which is less affected
by the fluctuation due to bright objects than the first version, for each
raw frame.
Thereafter the same procedures  were done to make the second-version
combined image. From second-version combined image, the mask image was
made again. Three iterations of these procedures were performed to
make final combined image. Figure \ref{HDFcisco} shows the combined
CISCO $K'$-band image of the HDF-N.
The FWHM of the point sources in 
the final combined image is 
about 0.55 arcsec. Since the field of view of the final image is
about $110'' \times 110''$, this $K'$-band image is not completely
covered the whole HDF-N area (Figure \ref{FOVcisco}).
Flux calibration to the Mauna Kea system $K$-band was carried out by
observing the United Kingdon Infrared Telescope (UKIRT) faint standard
stars \citep{haw01} taken before and after the HDF-N observation.

\subsection{HST WFPC2/NICMOS data}

  The HDF-N was observed with the HST NICMOS Camera 3
between UT 1998 June 13 and
June 23 (PI:M. Dickinson; PID 7817). The complete HDF-N was mosaiced
with eight sub fields in the $J_{110}$ and $H_{160}$-bands. We analyzed
the calibrated data of
these observations downloaded from the archival site of the Space
Telescope Science Institute. Each sub field was observed through nine dithered
pointings, with a net exposure of 12600s in each band, except for a few
pointings that could not be used due to a
guidance error of HST \citep{dic00b}.   We combined
these data into a single mosaiced image registered to the WFPC2
image of the HDF-N, using the ``drizzling'' method with the
$IRAF$ $DITHER$ package.

 For optical data, we analyzed the public ``version 2'' $U_{300}$, $B_{450}$,
$V_{606}$, $I_{814}$-band WFPC2 images of HDF-N produced by the STScI
team. These optical and near infrared HST
images were convolved with the Gaussian kernel 
to match the CISCO $K'$-band point-spread function for the aperture
photometry.

\subsection{Source Detection and Photometry \label{phot}}

 First, we performed source detection in the CISCO $K'$-band image
 using the SExtractor image analysis package \citep{ber96}. A detection
threshold of $\mu_{K}=22.91$ mag arcsec$^{-2}$, which corresponds to
about 1.5 times pixel-to-pixel variation of the background,
over 15 connected pixels was used. We adopted
MAG\_BEST from SExtractor as the total magnitude of each detected object. Of
the sources extracted by SExtractor, objects located at the edge of WFPC2
frames and those which we identified by eye as noise peaks (mostly near the
 bright objects) were rejected and removed from the final catalog.

\begin{figure}
\epsscale{0.8}
\plotone{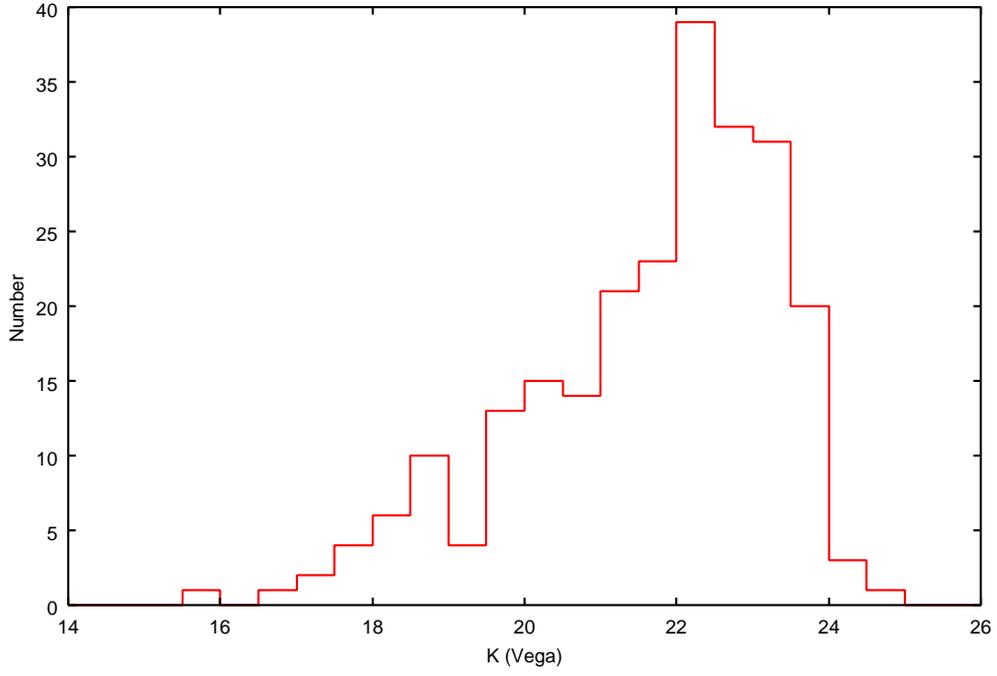}
\caption{CISCO $K'$-band number counts of the Hubble Deep Field North.
\label{NC}}
\end{figure}

 Figure \ref{NC} shows the raw number counts of the $K'$-band frame.
From this figure, we infer that the object detection is nearly complete to
$K\sim$22.0. In this source detection procedure, we could not find the
$K'$-selected object which was not detected in the HST $I_{814}$,
$J_{110}$, $H_{160}$-band images.

We cataloged the objects with $K<24.2$ (nearly all objects).
 To measure their color and SED, we selected the
aperture size at which S/N ratio at $K'$-band is 0.75 times
that at the maximum S/N aperture. This is compromise between
the highest S/N and measuring over the total lights of objects.
\begin{figure}
\epsscale{0.65}
\plotone{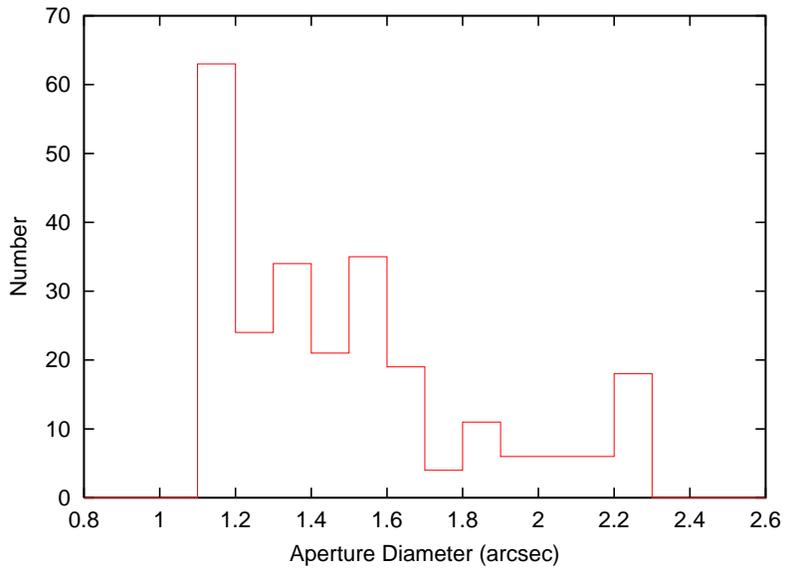}
\caption{Adopted aperture diameter distribution for seven optical--NIR
bands photometry.
\label{NCap}}
\end{figure}
The minimum aperture size is set to 1.1 arcsec diameter, which is
about two times larger than the seeing size of the images, to avoid
the seeing effect.
The adopted aperture size was between 1.1 and 2.3
arcsec in diameter, typically $\sim$1.5 arcsec (Figure \ref{NCap}).
With these apertures, we performed photometry for all seven 
$U_{300}$, $B_{450}$, $V_{606}$, $I_{814}$, $J_{110}$, $H_{160}$,
$K'$-band images. The background sky was estimated with the annulus,
which has 5 arcsec inner diameter and 1 arcsec 
width. The aperture position
for each object was fixed in order to measure the colors or broad-band
SED at the same physical region of the galaxy. 247 objects were cataloged
and their magnitudes were measured.

\subsection{Redshift Determination and Sample Selection}

 \citet{coh00} compiled the HDF-N galaxies with spectroscopic redshifts 
known by the time.
We identified the corresponding objects in our $K'$-selected 
catalogue by comparing the coordinates of the objects in Cohen et al.'s
catalogue and ours. A few additional spectroscopic redshifts from the 
Team Keck Treasury Redshift Survey \citep{wir04} also could be
identified in our catalogue. 
For 103 out of 247 objects, the
spectroscopic redshifts from the literature could be used.
Cohen et al.'s sample was selected mainly
by the standard $R \le 24$ mag, and
limited to $z \lesssim 1.3$, except for the Lyman
Break Galaxies at $z>2$ (e.g., \citealp{low97,dic98}).
TKRS has the similar selection criterion with slightly deeper magnitude
limit. 

 For those objects with no spectroscopic redshift, we estimated
 their redshifts
by using the photometric redshift technique with the photometric data
of the seven
optical--NIR bands mentioned in the previous subsection. We computed
the photometric
redshifts of those objects using the public code of $hyperz$ \citep{bol00}.
 The photometric redshift was calculated by $\chi^{2}$ minimization,
comparing the observed magnitudes with the values expected from a set of model
Spectral Energy Distribution. The free parameters involved in the
 fitting are the redshift, spectral type (star formation history), age, and
color excess (dust extinction).
We chose to use the SED model of Bruzual \& Charlot synthetic
library (GALAXEV; \citealp{bru03}) and the  Calzetti extinction law
\citep{cal00}.  

\begin{figure}
\epsscale{0.5}
\plotone{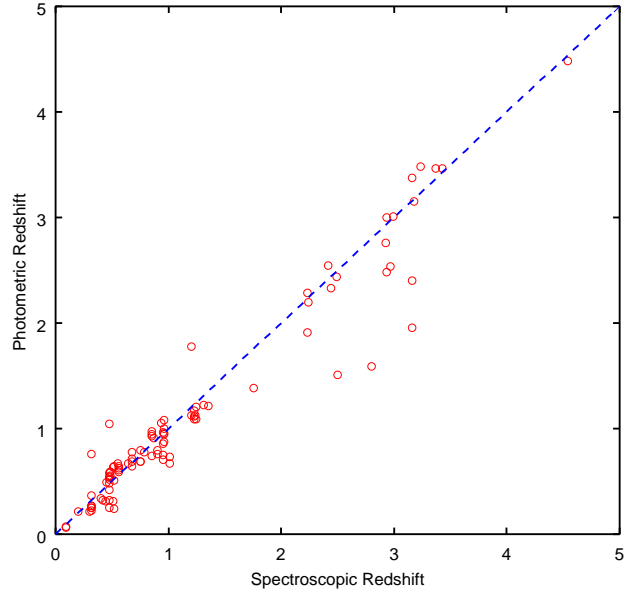}
\caption{Comparison between the spectroscopic
redshift and the photometric redshift
of the $K'$-selected galaxies with the spectroscopic redshift.
\label{testz}}
\end{figure}

 For those objects with spectroscopic redshifts,
we also computed the photometric
redshifts in order to test the accuracy of the photometric redshift
estimate for our
filter set. Figure \ref{testz} shows a comparison between the spectroscopic
redshifts and photometric redshifts for those objects in our catalogue.
The spectroscopic and the photometric redshifts agree well within
$\Delta z= 0.11$ over a wide redshift range, except for several
outliers. 

\begin{figure}
\epsscale{0.75}
\plotone{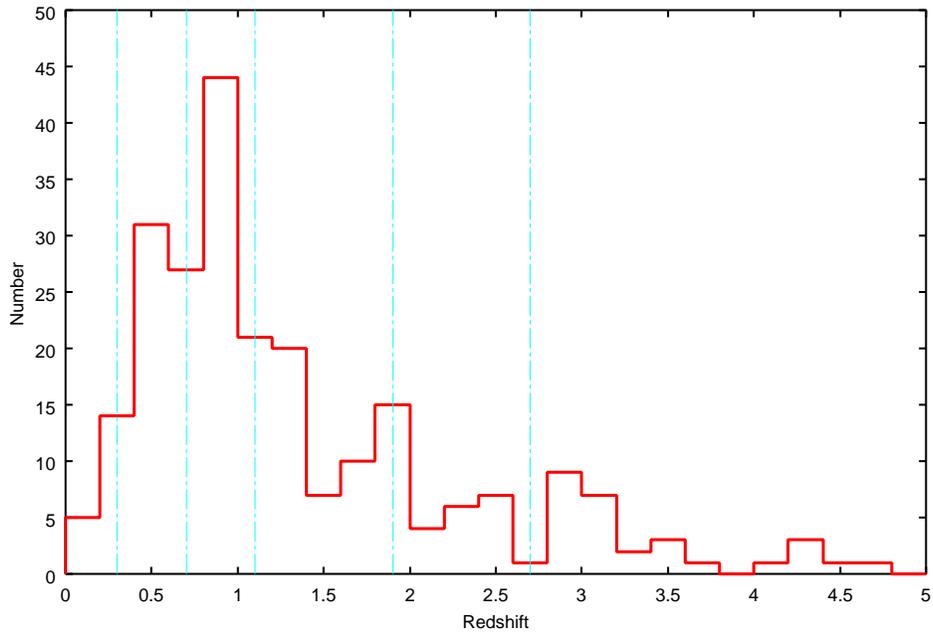}
\caption{Redshift distribution of our $K'$-selected
    galaxies. Vertical dotted-dash lines show the boundaries between the
    redshift bins used in the analysis.
\label{NCz}}
\end{figure}

Figure \ref{NCz} shows the redshift distribution of our
$K'$-band selected galaxies. 
We used the spectroscopic redshifts whenever possible, and 
adopted the photometric redshifts for the remainder.
We excluded those stars that 
were confirmed by spectroscopy
from our sample, and did not perform any further star/galaxy
 discrimination for fainter objects.
We use galaxies at $0.3<z<2.7$ in the following analysis.

\subsection{Stellar Mass and Rest-frame Color Estimation \label{smass}}

 In addition to photometric redshift estimation, using the seven
optical--NIR bands photometry, we performed the detailed SED fitting
to estimate the stellar mass and rest-frame color of our sample.
Like an above photometric redshift estimation, we used the GALAXEV 
synthetic library as the SED model.
The free parameters involved in the fitting are metallicity,
star formation timescale (see below), age, and
color excess. 
The reason for varying many parameters such as metallicity in this SED
fitting procedure is to investigate possible range of the stellar mass
of galaxies rather than to determine each free parameter precisely.
We assumed the exponentially decaying 
star formation rate with time (SFR $\propto e^{-t/\tau}$), and
varied the characteristic timescale
$\tau$ from 0.01 Gyr to 30 Gyr as a free parameter. Metallicity is
changed from 0.02 solar metallicity to 2.5 solar metallicity.
We assumed the \citet{cha03}'s
 initial mass function (IMF) with upper and lower mass cutoffs
m$_{l}=0.1$M$_{\odot}$ and m$_{u}=100$M$_{\odot}$, and the Calzetti
extinction law. 
The $\chi^{2}$ value for each GALAXEV template was
calculated, and the minimum $\chi^{2}$ value for each observed galaxy
was found. GALAXEV code outputs the stellar mass-to-light ratio (M/L)
for each template, and therefore we can estimate the stellar mass of 
objects from 
the total absolute magnitude and M/L. 
We calculated the (aperture) absolute $V$-band
magnitude M$_{V}$, using the best-fit SED template and the redshift
adopted above. The $U$-to-$K$ bands photometries allow us
to  measure the rest-frame $V$-band magnitude without extrapolation
(use only interpolation) for galaxies at $z \lesssim 2.8$.
For the objects only with photometric redshifts, we further varied redshift 
as a free parameter, and calculated the $\chi^{2}$ value for each 
GALAXEV template at various redshifts in order to investigate the effect
of the photometric redshift uncertainty on the stellar mass estimate.
We performed the conversion from the aperture magnitude to the total absolute
magnitude by using the differences between the aperture magnitude and 
the MAG\_BEST value at the observed $K$-band. 
Then the dust-extinction corrected absolute magnitude
was used to calculate the stellar mass. 
The cosmology of $H_{0}=70$ km s$^{-1}$ Mpc$^{-1}$,
$\Omega_{\rm 0}=0.3$, $\Omega_{\Lambda}=0.7$ is
assumed.

As well as absolute magnitude, the rest-frame color can be estimated from
the best-fit SED template. For analysis,
we use the rest $U-V$ color, which
is sensitive to the existence/absence of the largest
features in the galaxy continuum spectra at near-UV to optical wavelengths,
namely, a 4000 \AA\  break or the Balmer discontinuity.
Since these features are sensitive to the average age of stars
especially at relatively young stage, 
we can infer the evolution of the average
age of their stars even at the region where the photometric
uncertainty is relatively large.
The rest-frame $U-V$ color can be estimated without
extrapolation for objects at $z \lesssim 2.8$.

\subsection{Morphological Classification}

 We performed a quantitative morphological classification for the
$K'$-selected galaxies using the HST WFPC2/NICMOS
images, as described by \citet{kaj01}.
The classification method used is based on the central concentration
index, $C$, and the asymmetry index, $A$ \citep{abr96}.
The bands used for the classification were selected so that
the classification was made on the rest-frame $B$ to $V$-band
image for each object ($I_{814}$-band for $z<1.0$, $J_{110}$-band for
$1.0<z<2.0$, $H_{160}$-band for $z>2.0$), where various
 morphological studies of local galaxies have been done.
 Since the $C$ and $A$ indices depend on the brightness of the object,
even if the
intrinsic light profiles were identical, we evaluated the morphology 
by comparing the indices of the observed galaxy with those of a set of
simulated artificial ones that
have similar photometric parameters as each observed galaxy.
The detail processes of the classification are described in \citet{kaj01}.\\
Although some scatter exists,
our quantitative $C$/$A$ classification is confirmed to
correlate well with an eyeball classification
by \citet{vdb00} (\citealp{kaj01}, Figure 8).
The number of
galaxies which are classified into each morphological category is
40 in bulge-dominated, 23 in intermediate, 92 in disk-dominated, 65 in
irregular, respectively. The other 20 objects were cannot be
classified because of their faintness.

We prefer to use a quantitative morphological classification, because
it is not only more objective, but can also take into account
the effect of the magnitude/surface brightness bias for classification.

\section{Results}
 Using the very deep $U$-to-$K$ band images of the Hubble Deep Field
North, we can estimate the rest-frame optical properties of galaxies
to $z \sim 3$, and investigate the distribution of the stellar mass,
$U-V$ color, morphology.

\subsection{Stellar Mass Distribution}

\begin{figure}
\epsscale{0.9}
\plotone{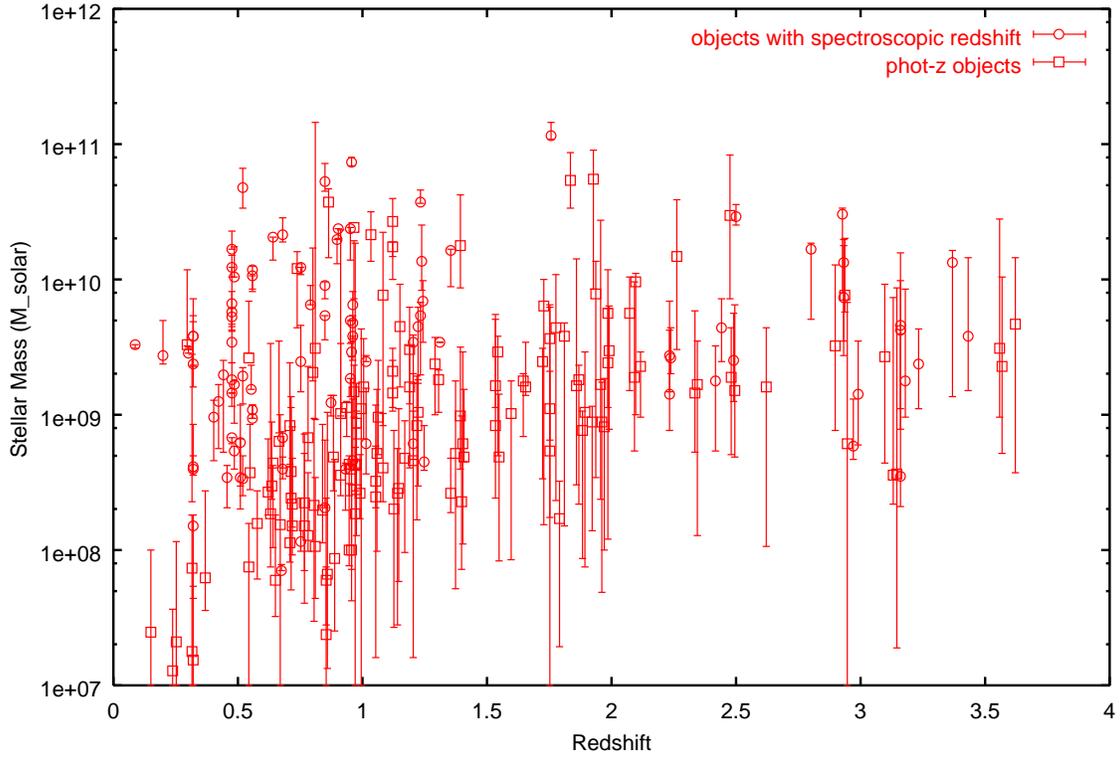}
\caption{Stellar mass distribution of our $K'$-selected sample
of galaxies in the HDF-N as a function of redshift. 
Circles show objects with spectroscopic redshifts and squares 
represent the phot-z sample.
Error bars show the range of 90\% confidence level.
\label{Ms_z}}
\end{figure}

 Figure \ref{Ms_z} shows our estimate of the stellar mass of the sample
 as a function of redshift. Error bars indicate the range of
90\% confidence level, which is estimated from the $\chi^{2}$ value 
at each position in the parameter space in the SED fitting.
Circles show the objects with spectroscopic redshifts and 
squares represent the phot-z sample. For the phot-z sample, 
we took account of the uncertainty of the photometric redshift 
in the error estimate. 
Because our sample is limited to $K<24.2$, 
only the more massive galaxies tend to be selected at the higher redshift. 
While we can sample galaxies to the stellar mass of
$\sim1\times10^{8}$M$_{\odot}$ at $z\sim1$, only galaxies with the stellar
mass larger than $\sim 1\times10^{9}$M$_{\odot}$ can be 
detected at $z\sim2$. Less massive,
higher redshift galaxies have fainter apparent magnitude, and therefore the
constraint on the stellar mass of the galaxies tends to be weak.
Furthermore, the photometries only at 
shorter rest-frame wavelength can be used for galaxies at higher redshift,
 which weakens the constraint 
on the stellar mass. For example, the typical 1$\sigma$ error of the 
mass estimate for the galaxies with stellar mass of
$10^{9}-10^{10}$M$_{\odot}$ is $\sim0.2$ dex at 
$z\sim1$, and becomes $\sim0.4$ dex at $z\sim2$.
\citet{dic03} estimated the stellar mass
of H$_{160}$-selected galaxies in the HDF-N, using the HST
WFPC2/NICMOS images and the ground-based $K$-band data with IRIM on
the KPNO 4 m.
Our result about the stellar mass distribution 
 is similar with that in Dickinson et al., although the selection
and the field of view are slightly different. 
Our stellar mass estimate with \citet{cha03}'s
IMF is about 1.8 times smaller systematically than that with Salpeter IMF,  
which is adopted in \citet{dic03}.

In Figure \ref{Ms_z}, it is seen that there is few galaxies with the
stellar mass larger than 1 $\times 10^{11}$ M$_{\odot}$.
With small corresponding volume of the HDF-N, the expected number of
galaxies with M$_{stellar} > 1\times 10^{11}$ M$_{\odot}$ is 
about one at $z\lesssim1$,  
which is calculated from the results of 2MASS and 2df survey for local
galaxies \citep{col01} assuming no evolution.
At $1<z<3$, the expected number of these galaxies becomes about nine,
while we detected only one galaxy with  M$_{stellar} > 1\times
10^{11}$ M$_{\odot}$ in our sample. 
The number density of these massive galaxies seems to decrease at
$z\gtrsim1$ in the HDF-N, but the significance 
 of such small number statistics may be relatively low, 
considering the field-to-field variance due to the clustering of
galaxies. To investigate the number density, color, and morphology
distribution of these massive galaxies with a stellar mass larger than
$10^{11}$ M$_{\odot}$, the larger-volume survey is needed.

\subsection{Rest $U-V$ Color vs Stellar Mass}

\begin{figure}
\epsscale{0.65}
\plotone{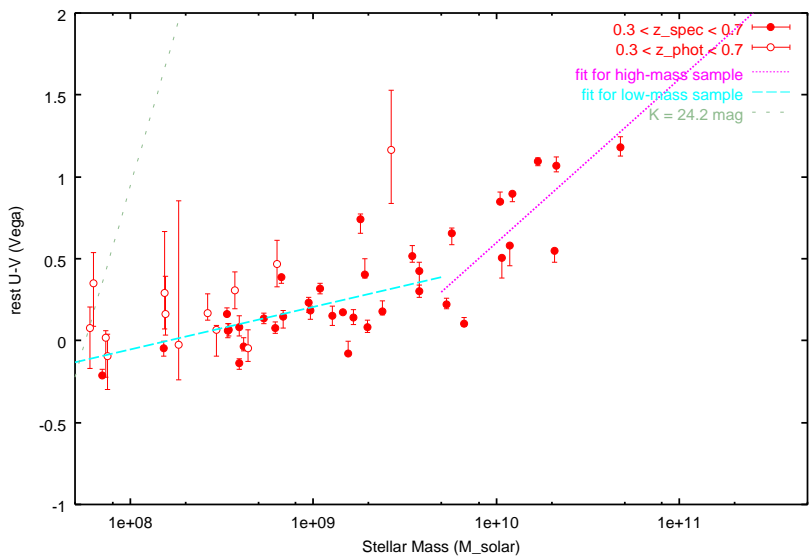}
\caption{rest frame $U-V$ color distribution of galaxies at
    $0.3<z<0.7$ as a function of stellar mass.
    Solid circles represent the objects with spectroscopic redshifts,
    and open circles show the phot-z sample. 
    Error bars represent the range of 90\% confidence level.
    Short-dashed line shows the corresponding 
    detection limit estimated from the various GALAXEV models with $K=24.2$.
    Dotted line and long-dashed line represent the fitting results for 
the high-mass sample and the low-mass sample.
\label{Ms_UV03}}
\end{figure}
\begin{figure}
\epsscale{0.65}
\plotone{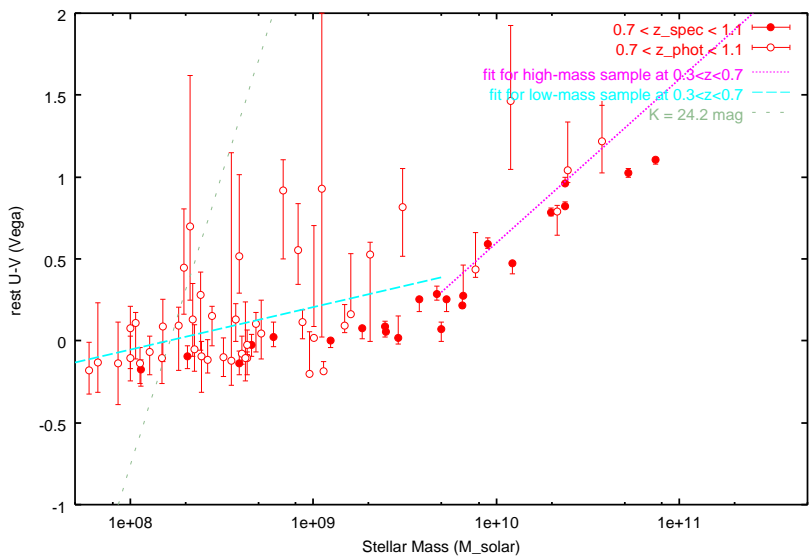}
\caption{Same as Figure \ref{Ms_UV03}, but
for galaxies at $0.7<z<1.1$. For comparison, the fitting results for
the low-mass sample and high-mass sample at $0.3<z<0.7$ are showed.
\label{Ms_UV07}}
\end{figure}
\begin{figure}
\epsscale{0.65}
\plotone{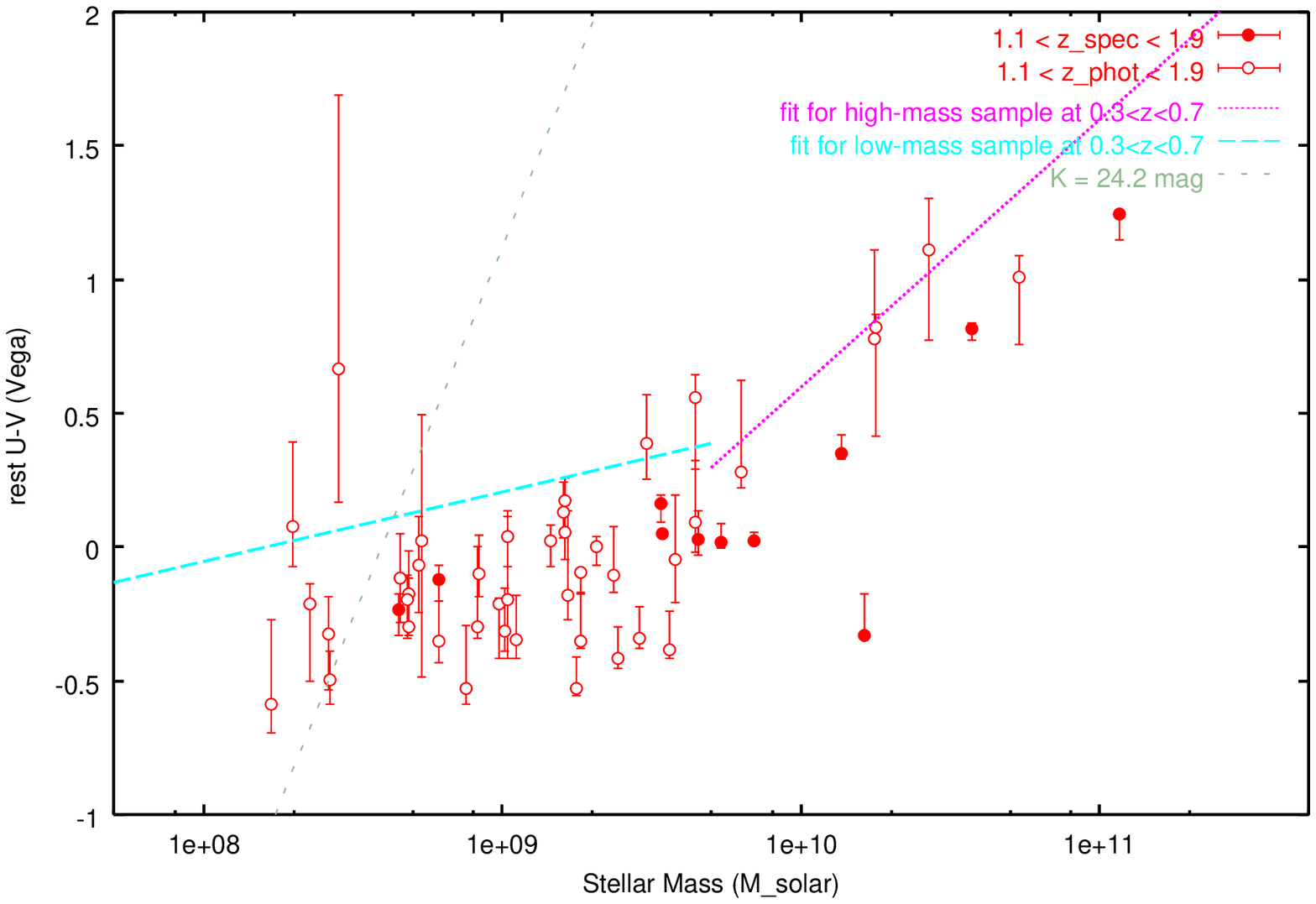}
\caption{Same as Figure \ref{Ms_UV07}, but
for galaxies at $1.1<z<1.9$.
\label{Ms_UV1}}
\end{figure}
\begin{figure}
\epsscale{0.65}
\plotone{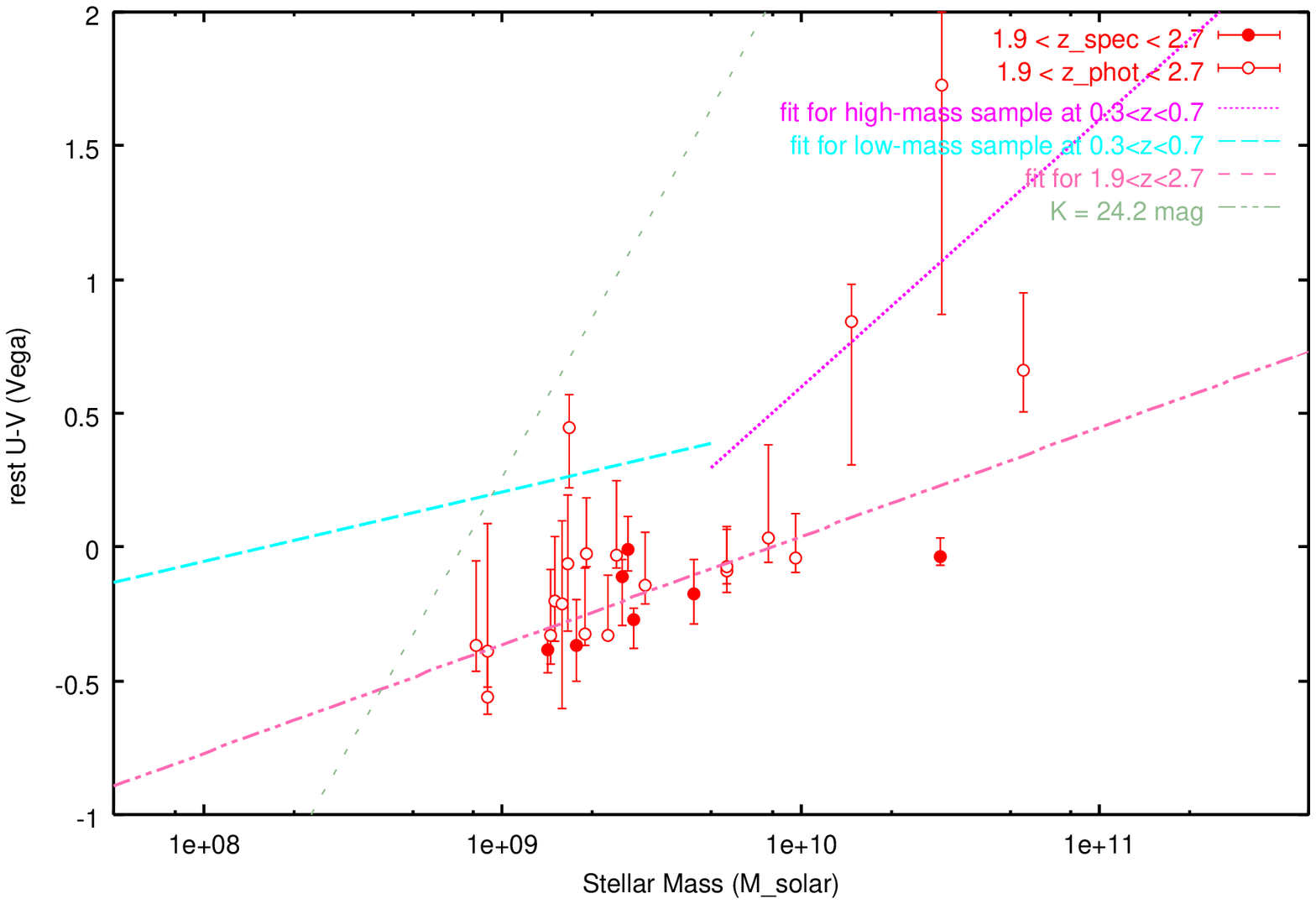}
\caption{Same as Figure \ref{Ms_UV07}, but
for galaxies at $1.9<z<2.7$. 
\label{Ms_UV2}}
\end{figure}
Figure \ref{Ms_UV03}-\ref{Ms_UV2} show the rest $U-V$ color
 distribution of our $K'$-selected galaxies in the HDF-N as 
a function of stellar mass for each redshift bin. 
For local galaxies, 
the $U-V$ color range is $\sim1.2$-1.5 for E and S0 galaxies, 
$\sim0.7$-1.1 for Sa-Sb galaxies, and $\sim0$-0.6 for Sc-Irr galaxies, 
respectively.\\
The range of each redshift bin is selected such that the 
photometric redshift uncertainty does not affect strongly
the membership in each redshift bin. In figure \ref{NCz}, which shows 
the redshift distribution of our sample, vertical dotted-dash lines
represent the boundary of each redshift bin.
In our assumed $H_0$ = 70 km s$^{-1}$ Mpc$^{-1}$, $\Omega_{\rm 0}=0.3$,
$\Omega_{\Lambda}=0.7$ cosmology,
the corresponding co-moving volume of each redshift bin is
1332 Mpc$^{3}$ for $0.3<z<0.7$ bin, 2720 Mpc$^{3}$ for $0.7<z<1.1$,
7885 Mpc$^{3}$ for $1.1<z<1.9$ bin, 8948 Mpc$^{3}$ for $1.9<z<2.7$,
respectively.
The number of objects in each redshift bin is 53 in $0.3<z<0.7$ bin,
70 in $0.7<z<1.1$ bin, 54 in $1.1<z<1.9$ bin, 28 in $1.9<z<2.7$,
respectively. 
Solid symbols represent the objects with spectroscopic
 redshifts and open symbols show those with photometric redshifts. 
Error bars in Figure \ref{Ms_UV03}-\ref{Ms_UV2}
show the $U-V$ range of 90\% confidence level for the SED fitting.
For the phot-z sample, the uncertainty of photometric redshift is taken 
into account in the confidence level.\\
The detection limit corresponding to $K=24.2$ was estimated from the 
GALAXEV models. Using the GALAXEV models with various age and star
 formation timescale $\tau$ (solar metallicity and no dust extinction
 are assumed), we calculated the stellar masses and the rest $U-V$ colors
 for the models with $K=24.2$ and the corresponding redshift range.
 We simply performed the linear fit 
 for these calculated masses and colors,  
 and the result is showed as the short-dashed line in 
Figure \ref{Ms_UV03}-\ref{Ms_UV2}.

In Figure \ref{Ms_UV03} and \ref{Ms_UV07}, 
it is seen that there is few galaxies with
relatively blue color (e.g., $U-V<0.3$) at M$_{stellar} >
1\times10^{10}$M$_{\odot}$. On the other hand, at M$_{stellar}
\lesssim 5\times10^{9}$M$_{\odot}$, the number of objects with
red color (e.g., $U-V>0.6$) is much smaller than the bluer galaxies.
At M$_{stellar} \sim 5\times10^{9}$M$_{\odot}$, the transition of
rest $U-V$ color distribution seems to occur.
In Figure \ref{Ms_UV1}, the similar transition of the color is also 
seen at $1.1<z<1.9$. At M$_{stellar}\lesssim5\times10^{9}$M$_{\odot}$, 
most galaxies have $U-V\lesssim0$, while redder galaxies dominate at 
 M$_{stellar}\gtrsim5\times10^{9}$M$_{\odot}$.
At $1.9<z<2.7$ (Figure \ref{Ms_UV2}), the number of red (e.g.,
$U-V\gtrsim0.5$) or massive (e.g., $\gtrsim10^{10}$M$_{\odot}$) 
galaxies is very small, and it is uncertain whether the trend seen at 
lower redshifts exists or not.

\begin{figure}
\epsscale{0.75}
\plotone{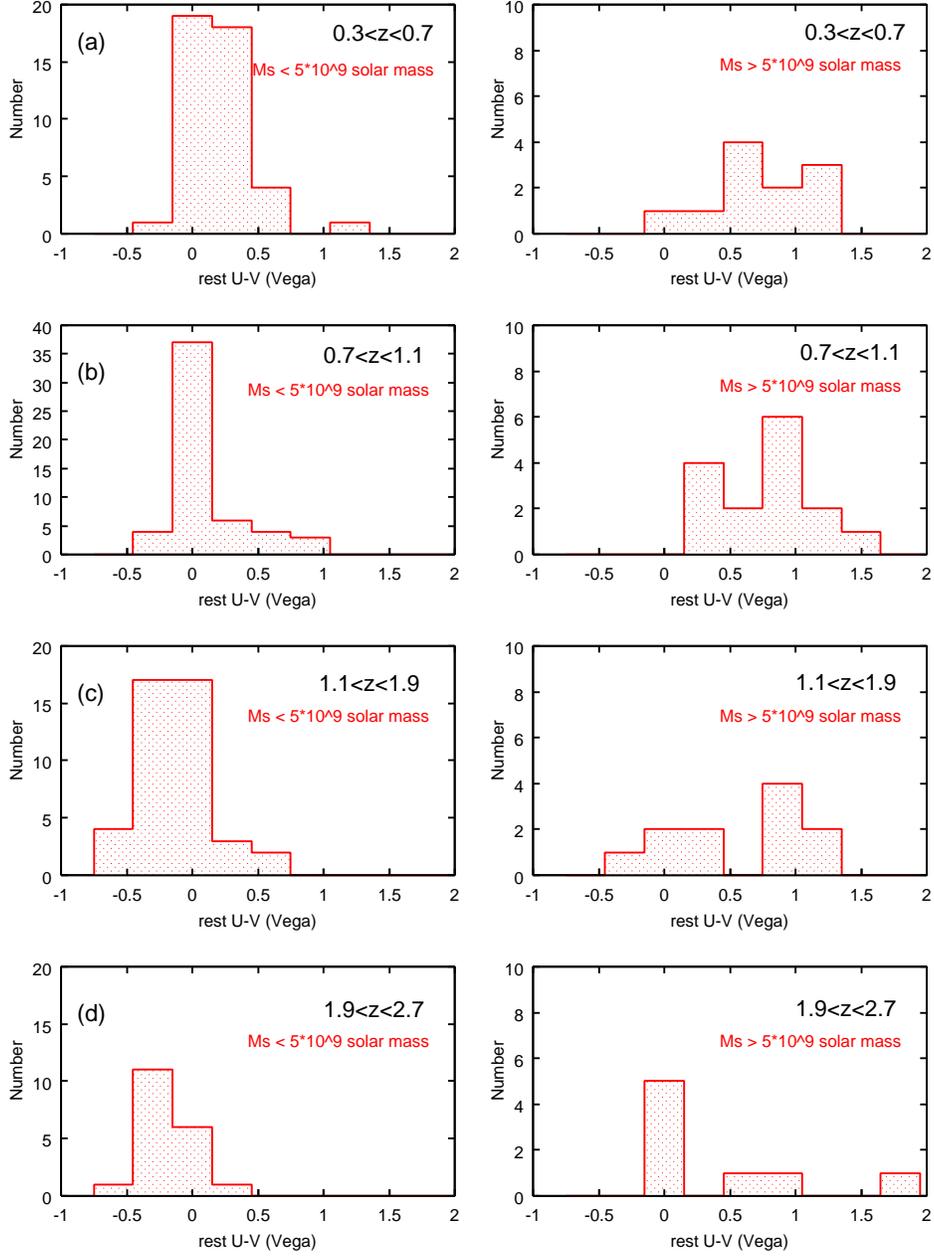}
\caption{Rest $U-V$ color distribution of galaxies 
when the sample is divided
into two populations at 5$\times10^{9}$M$_{\odot}$. left:for galaxies
    with M$_{stellar}<5\times10^{9}$M$_{\odot}$, right:for galaxies
with M$_{stellar}>5\times10^{9}$M$_{\odot}$. Each row corresponds to
each redshift bin. (a):$0.3<z<0.7$, (b):$0.7<z<1.1$, (c):$1.1<z<1.9$, 
(d):$1.9<z<2.7$.
\label{UVhist}}
\end{figure}

To contrast the transition of the rest-frame color, in Figure \ref{UVhist}, 
we show the $U-V$ color histogram
for galaxies with M$_{stellar}<5\times10^{9}$M$_{\odot}$
and  M$_{stellar}>5\times10^{9}$M$_{\odot}$, respectively.
In the Figure \ref{UVhist}a and \ref{UVhist}b, 
the differences of 
the $U-V$ distributions of these low- and high-mass samples are seen.
While the rest $U-V$ color distribution of the low-mass sample shows
the peak at $U-V\sim$0-0.3, the high-mass sample has redder $U-V$ distribution.
In figure \ref{UVhist}c and \ref{UVhist}d, 
we show the same comparison 
of rest $U-V$ color distribution for galaxies at $1.1<z<1.9$ and
$1.9<z<2.7$.  At $1.1<z<1.9$, the similar trend with that at $z<1.1$
is clearly seen.
On the other hand, this trend is less secure at $1.9<z<2.7$.

In Figure \ref{Ms_UV03} and \ref{Ms_UV07}, 
the low- and high-mass samples seem to
have different dependences of rest $U-V$ color on their stellar
mass. While the rest $U-V$ color of the high-mass galaxies is strongly
correlated with the stellar mass (more massive galaxies have
redder color), the correlation between the $U-V$ color and stellar mass
seems to be very weak in the low-mass population.
To quantify this trend, we performed the linear fit in the
rest $U-V$ vs $\log{M_{stellar}}$ plane for the high-mass and the low-mass
sample, respectively at each redshift bin. 
The result is summarized in Table \ref{slope}.
The estimated value of slope is
0.260$\pm$0.096 for the low-mass population, 1.002$\pm$0.269 for 
the high-mass
population at $0.3<z<0.7$.  At $0.7<z<1.1$, the slope for the low-mass
sample is 0.152$\pm$0.095, and that for the high-mass sample is 0.801$\pm$
0.196. These results confirm that the rest $U-V$ color of  galaxies with
M$_{stellar}<5\times10^{9}$M$_{\odot}$ is only weakly
correlated (or not correlated) with stellar mass.
At $1.1<z<1.9$, the fitted slope is  0.049$\pm$0.119 for the low-mass
sample, 0.980$\pm$0.279 for the high-mass sample, which shows that the
similar trend with that at $z\lesssim 1$ is maintained at this redshift range.
Results for galaxies at $1.9<z<2.8$ is 0.450$\pm$0.546 for the low-mass
sample, 0.425$\pm$0.437 for the high-mass one. We cannot find
the difference of the $U-V$ color dependence on stellar mass
between the samples divided at 5$\times10^{9}$M$_{\odot}$ at this
redshift range, as can be seen from Figure \ref{Ms_UV2}.
Instead, over the mass range of
1$\times10^{9}$--1$\times10^{10}$M$_{\odot}$, the trend that galaxies
with higher stellar mass have redder $U-V$ color is seen.
We performed the same linear fitting as above for all (low-mass and
high-mass samples) galaxies in $1.9<z<2.7$ bin, and showed the result
in Figure \ref{Ms_UV2}. The estimated slope is 0.405$\pm$0.162.

It should be noted that our $K'$-selected sample does not strictly
correspond to the stellar mass-selected one. 
If we consider the same stellar mass
galaxies at the same redshift, 
galaxies with the redder color tend to have the lower $K$-band magnitude, 
because they are dominated by older stellar population or are highly
dust-extincted, and have higher mass-to-light ratio.
Therefore, near the lower limit of stellar mass of the sample
(near the detection limit of our $K'$-band observation),
there is the bias against the detection of redder galaxies in each
redshift bin. At least $z\lesssim2$, however, the rest $U-V$
distribution seen in Figure \ref{Ms_UV03}-\ref{Ms_UV1} 
become rather blue and the stellar
mass dependence of the $U-V$ color becomes weak at the stellar
mass of $\sim5\times10^{9}$M$_{\odot}$, which corresponds to
the value much higher than the detection limit of our observation,
and therefore this bias does not seem to 
influence significantly our results about the color
distribution and their dependence on stellar mass.\\
At $1.9<z<2.7$, where the $K$-band detection limit corresponds to
$\sim1\times10^{9}$M$_{\odot}$, the bias against the redder galaxies may 
affect the slope of the correlation between the stellar mass and the
$U-V$ color (see short-dashed line in Figure \ref{Ms_UV2}). 
As seen in Figure \ref{Ms_UV2}, however, there is the trend
that the $U-V$ color of the bluest galaxies at each stellar mass
becomes bluer as the stellar mass decreases, 
which cannot be explained by this bias.\\
Furthermore, in order to investigate the stellar mass value where the
change of the mass dependence of the rest $U-V$ color occurs, we retried
the linear fit in the rest $U-V$ vs $\log{M_{stellar}}$ plane, varying
the boundary mass between the high- and low-mass samples as a free
parameter in each redshift bin. The best fit boundary stellar mass
is 6.76$^{+5.24}_{-3.29} \times 10^{9}$M$_{\odot}$ at $0.3<z<0.7$,
7.08$^{+12.9}_{-5.26} \times 10^{9}$M$_{\odot}$ at $0.7<z<1.1$,
3.63$^{+13.0}_{-2.19} \times 10^{9}$M$_{\odot}$ at $1.1<z<1.9$, respectively.
We could not find the significant
evolution of the stellar mass where the mass dependence of the 
rest $U-V$ color changes, although we could not constrain strongly
 this stellar mass in each redshift bin.

To estimate the degree of the evolution of the $U-V$ color distribution
as a function of stellar mass, in Figure
\ref{Ms_UV07}-\ref{Ms_UV2}, 
we compared the
$U-V$ distribution of galaxies at $0.7<z<1.1$, $1.1<z<1.9$,
$1.9<z<2.7$, respectively, with the fitting result for those at
$0.3<z<0.7$. From these figures, it is seen that at M$_{stellar} <
5\times10^{9}$M$_{\odot}$, the rest $U-V$ distribution become
gradually bluer with redshift. At $0.3<z<0.7$, the envelope of
galaxies with the bluest $U-V$ color lies at $U-V\sim-0.1$.
On the other hand, the bluest galaxies at $1.1<z<1.9$ have $U-V\sim -0.5$.
For simplicity, we ignored the small slope
in the $U-V$ vs stellar mass plane for the low-mass sample and calculated the
average $U-V$ value of galaxies with
M$_{stellar}<5\times10^{9}$M$_{\odot}$ at each redshift bin.
The result is 0.153$\pm$0.173 at $0.3<z<0.7$, 0.019$\pm$0.142 at
$0.7<z<1.1$, -0.081$\pm$0.157 at $1.1<z<1.9$, -0.241$\pm$0.186 at
$1.9<z<2.7$. \\
On the contrary, the relation between the stellar mass and rest
$U-V$ color of galaxies with M$_{stellar}>5\times10^{9}$M$_{\odot}$
does not seem to change significantly at $z\lesssim2$. In fact,
the rest $U-V$ distribution of the high-mass sample at $0.7<z<1.1$ and
$1.1<z<1.9$ corresponds well to the fit for the high-mass galaxies at
$0.3<z<0.7$, in Figure \ref{Ms_UV07} and \ref{Ms_UV1}. 
We performed the fitting for
 the high-mass sample at $0.7<z<1.1$ and $1.1<z<1.9$, assuming the
same slope as that at $0.3<z<0.7$, and compared the resultant
intercept value with that at  $0.3<z<0.7$. The difference of intercept
is -0.115$\pm$0.274 at $0.7<z<1.1$, -0.224$\pm$0.362 at $1.1<z<1.9$,
respectively, which shows the differences are less significant.
At $1.9<z<2.7$, the number of
galaxies with M$_{stellar}>1\times10^{10}$M$_{\odot}$ is very small, 
and the relation between the stellar mass and the rest $U-V$ seems to
disappear. These results about the degree of the evolution of the rest
$U-V$ color are summarized in Table \ref{cept}.

\begin{figure}
\epsscale{0.55}
\plotone{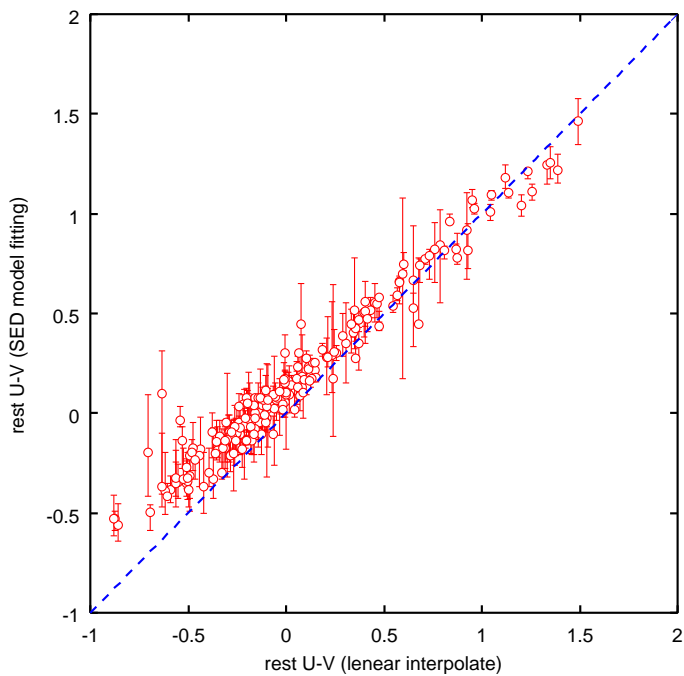}
\caption{Comparison between the rest-frame $U-V$ colors estimated by
    different procedures(linear interpolation vs SED fitting with
    population synthesis model). Error bars represent the range of
90\% confidence level.
\label{testcolor}}
\end{figure}

The results obtained in this section are about the relation between
the stellar mass and the rest-frame $U-V$ color of galaxies. Since
both the stellar mass and the rest $U-V$ were calculated from the SED
fitting with the population synthesis model, there can be possibility that
some biases about the SED model work on the results and produce the
artifacts.
To check this, we estimated the rest $U-V$ color without the SED fitting
procedure, using the direct linear interpolation from the observed
photometries. Figure \ref{testcolor} 
shows the comparison between the rest $U-V$
colors estimated by the SED fitting with GALAXEV model and the linear
interpolation. Although the estimation with SED fitting tend to
over/under-estimate the rest $U-V$ color at very blue/red region, the
correspondence between these two estimations is good, and the results
obtained above do not change if we use the $U-V$ value
calculated from the linear interpolation.

\begin{figure}
\epsscale{0.45}
\plotone{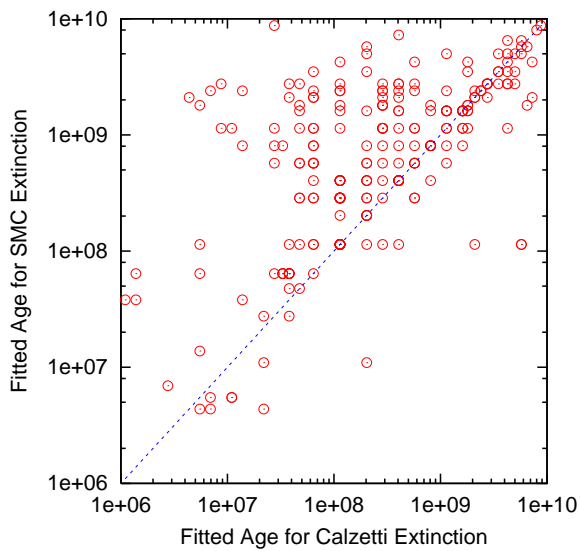}
\caption{Comparison between the fitted ages estimated by the SED fitting
   with  different extinction curves(Calzetti law vs SMC like extinction)
\label{testage}}
\end{figure}
\citet{fon03} pointed out that the SED fitting with
the Calzetti's extinction law yields lower fitted age than that with
the Small Magellanic Cloud (SMC) like extinction curve, and that the
stellar mass estimates with Calzetti law are slightly smaller.
In order to 
check the uncertainty due to the differences of adopted extinction curves, 
we performed the SED fitting with SMC extinction
\begin{figure}
\epsscale{0.55}
\plotone{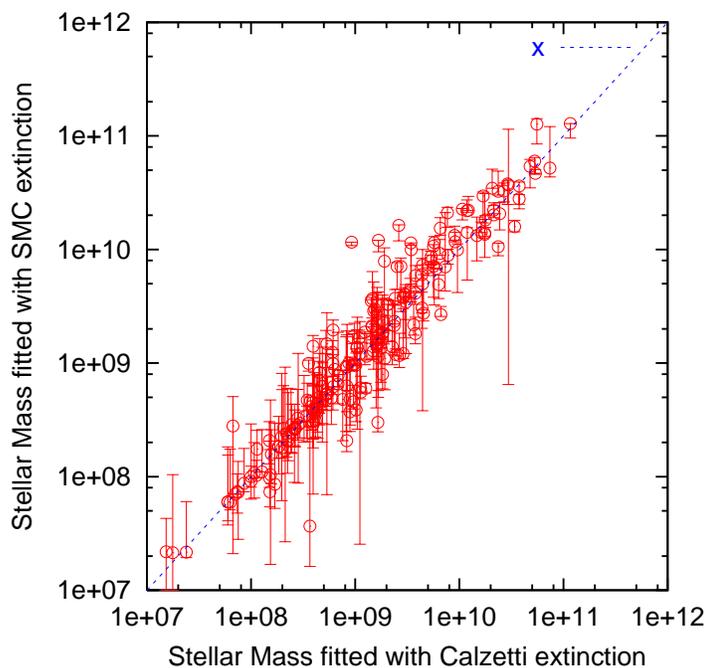}
\caption{Comparison between the stellar masses estimated by the SED fitting
   with  different extinction curves(Calzetti law vs SMC like
extinction). Error bars represent the range of 90\% confidence level.
\label{testmass1}}
\end{figure}
\begin{figure}
\epsscale{0.6}
\plotone{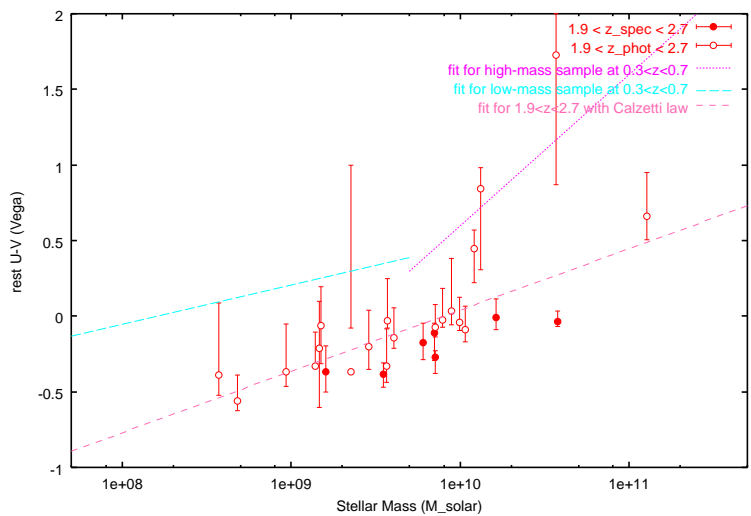}
\caption{same as Figure \ref{Ms_UV2}, 
except that the SMC extinction curve is used in
    the SED fitting for the stellar mass estimation.
\label{testMsUV1}}
\end{figure}
law \citep{pei92}, where other factors are not changed from the fitting
with Calzetti law. Figure \ref{testage}
 shows the comparison of the fitted ages
estimated with different extinction laws. As pointed out by \citet{fon03}, 
the ages estimated with SMC extinction clearly tend to have higher
value than that with Calzetti law. 
We calculated 
the ratio of the stellar mass estimated with SMC extinction relative to
that with Calzetti law, and found that the average of the ratio of estimated
stellar masses is  0.99$\pm0.05$.
As showed in Figure \ref{testmass1}, 
the stellar masses estimated from the fitting with different extinction
laws seem to agree well with a few exceptions.
Since
the fit with SMC extinction yields higher age, estimated mass-to-luminosity
ratios tend to be higher. But the degree of attenuation by dust tends
to be smaller value in the fit with SMC extinction, and the estimated
dust-corrected absolute magnitudes are fainter than that with Calzetti
law. Because these effects tend to be canceled out, the differences in
the stellar mass become smaller, especially in the case that the
photometries at the long-ward wavelength constrains the rest-frame NIR
flux strongly. Most of the outliers seen in Figure \ref{testmass1} 
are galaxies in
the $1.9<z<2.7$ bin. We recalculated the ratio of the stellar masses 
of galaxies at $1.9<z<2.7$ estimated with different extinction laws,
and the result is 1.41$\pm0.17$. For these galaxies, whose observed
$K$-band photometries reach only the rest-frame $V$ to $R$-band,
there seems to be the trend indicated by \citet{fon03} that the
SED fitting with SMC extinction yields slightly larger stellar mass.
For reference, we show the rest $U-V$ color distribution of galaxies at
$1.9<z<2.7$ as a function of the stellar mass estimated with the SMC
extinction in Figure \ref{testMsUV1}. Overall trend is not changed.
\begin{figure}
\epsscale{0.55}
\plotone{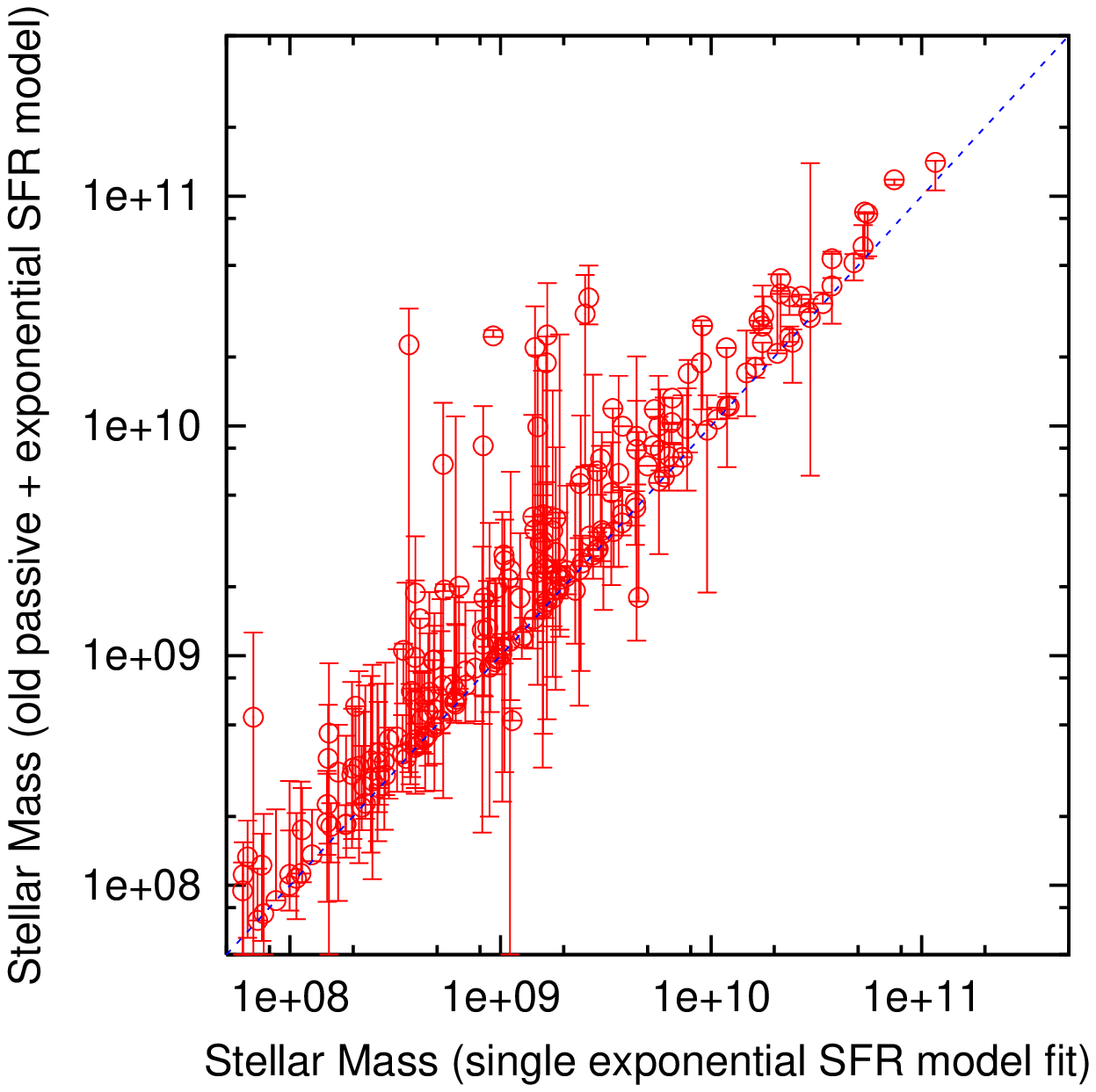}
\caption{Comparison between the stellar masses estimated by the SED fitting
   with different star formation history models 
(single exponentially decaying SFR model vs old passive population $+$
    exponential SFR model). Error bars show the range of 90\%
confidence level.
\label{testmass2}}
\end{figure}
\begin{figure}
\epsscale{0.6}
\plotone{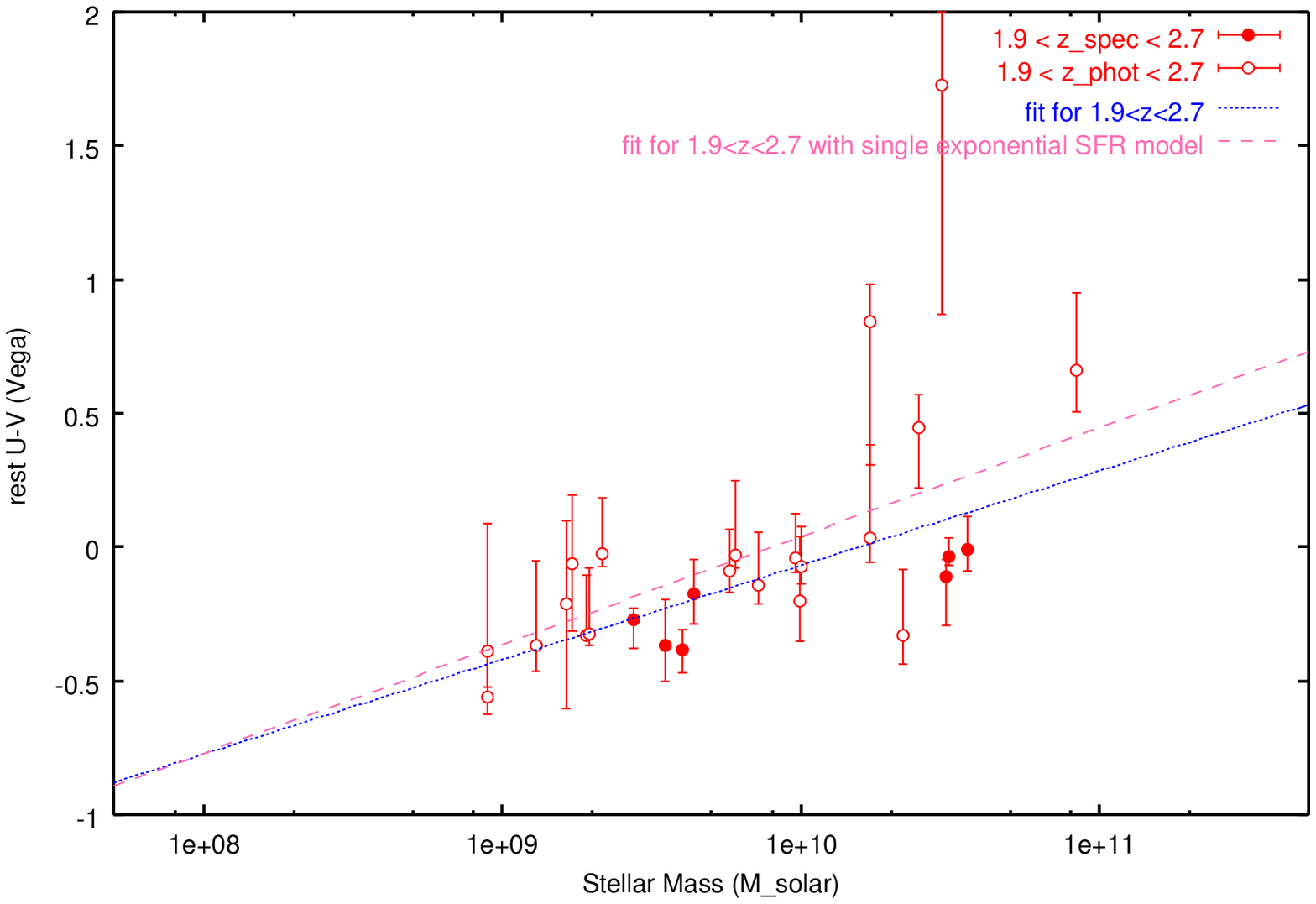}
\caption{same as Figure \ref{Ms_UV2}, except that 
the old passive population $+$
    exponential SFR model is used in
    the SED fitting for the stellar mass estimation.
\label{testMsUV2}}
\end{figure}

Although 
we assumed simple exponentially decaying star formation rate model
(characteristic timescale $\tau$ is free parameter) 
in the SED fitting procedure as mentioned above,
we also tried to use two-component star formation history model,
exponentially decaying SFR $+$ old passive-evolving population,
following \citet{pap01}, in order to estimate the upper limit of the
stellar mass of our sample. For the old population, 
we chose the single 0.01 Gyr burst model (after the burst, it
evolves passively) and fix its age to 
the age of the Universe at the observed redshifts
(formation redshift is infinity).
Maximum age for the old passive population yields maximum
mass-to-light ratio, and is suitable for the estimate of the upper
limit mass.
The relative
flux ratio between exponential SFR model and old population is added
as a free parameter to the SED fitting. The result is showed in Figure 
\ref{testmass2}. The distribution of the
best-fit stellar mass estimated with old population $+$
exponential SFR model is systematically higher than that with single
exponentially decaying SFR model. The average of the ratio of those
estimated with different star formation history models is 1.26$\pm0.05$.
The average ratio of the 68\% upper mass limits from the
two-component model relative to the best-fit stellar masses from
single exponential SFR model is $2.30\pm0.60$.
Again, most of outliers in Figure \ref{testmass2}
 are galaxies in the $1.9<z<2.7$
bin. At $z\lesssim2$, if we use the best-fit stellar mass estimated with
two-component star formation history model, the overall trend about the
distribution of $U-V$ color and the stellar mass is not
changed. At $1.9<z<2.7$, the average  ratio of the best-fit stellar
masses estimated with different star formation history models becomes
$1.54\pm0.21$, and those of the 68\% upper mass limits is
6.84$\pm1.31$.
For these galaxies, their faintness and relatively poor rest-frame
wavelength coverage cause large uncertainty. As seen in Figure \ref{testMsUV2},
however, the trend that galaxies with the higher stellar mass have
the redder rest $U-V$ color is still seen, even if we use the stellar
mass estimated with old population $+$ exponential SFR model.
The slope of the correlation between the $U-V$ color and the stellar
mass becomes $0.391\pm0.196$.

Furthermore, in order to check the effect of our adopted aperture size 
(section \ref{phot}, Figure
\ref{NCap}), we performed  
the fixed aperture photometry with 1.2, 2.4, 4.0 arcsec diameter
aperture, and calculated the rest $U-V$ color with the same way for
\begin{figure}
\epsscale{0.6}
\plotone{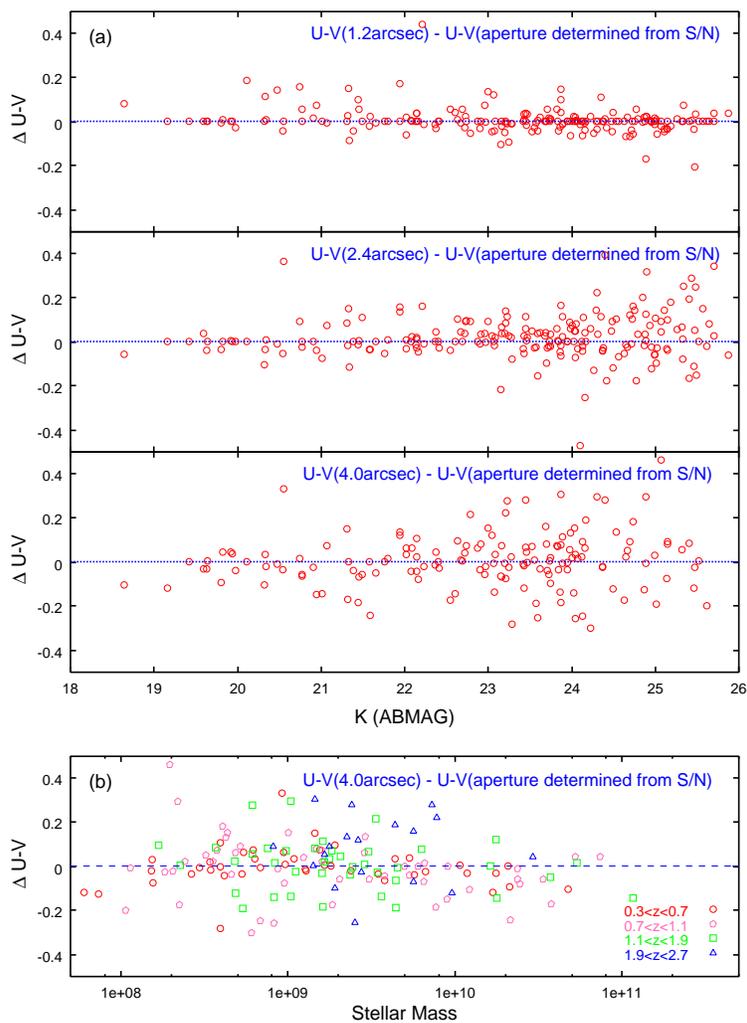}
\caption{Differences between the rest $U-V$ colors with the apertures 
determined from S/N ratio and those with fixed apertures. (a):
differences of $U-V$ as a function of $K$-band magnitude. top: for 1.2
arcsec diameter aperture. middle: for 2.4 arcsec diameter
aperture. bottom: for 4.0 arcsec diameter aperture. (b): differences of
$U-V$ as a function of stellar mass for 4.0 arcsec
diameter aperture. Each symbol represents each redshift bin. 
\label{UVap}}
\end{figure}
each aperture. The differences between the $U-V$ colors with our
adopted aperture and those with fixed apertures are showed in Figure
\ref{UVap} as a function of $K$-band magnitude and  stellar mass.
From the Figure \ref{UVap}(a), 
it is seen that for some bright galaxies, rest $U-V$ colors become
bluer as the aperture size increases. However the differences from the
 originally adopted aperture are at most $\sim0.2$ mag even for a 
very large 4.0 arcsec diameter aperture. Further, in Figure
\ref{UVap}(b), which show the aperture effects as a function of 
stellar mass, no systematic effect is seen. \\
The aperture effect  cannot
change our results about $U-V$ color distribution of galaxies 
as a function of stellar mass.

\begin{figure}
\epsscale{0.6}
\plotone{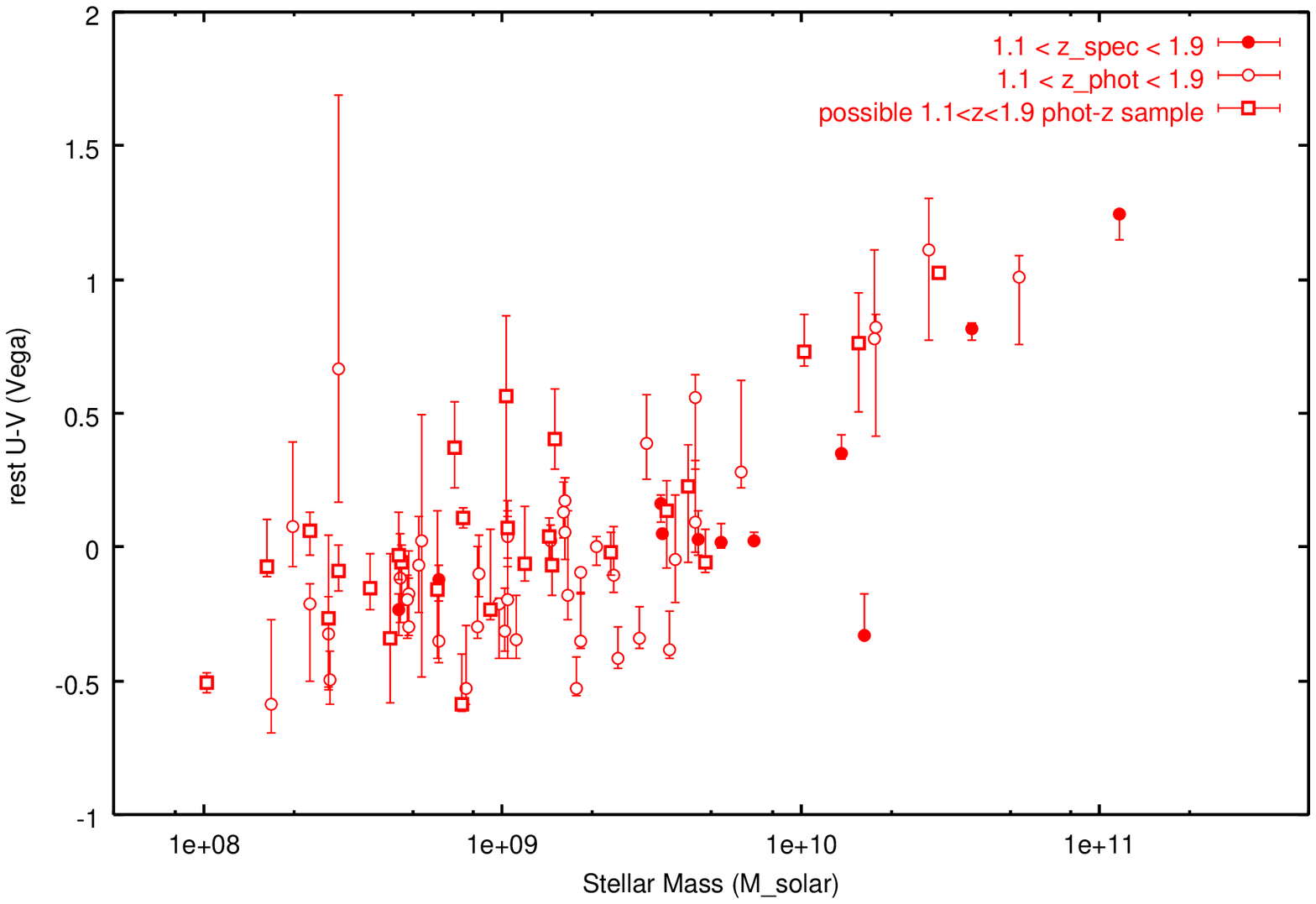}
\caption{Same as Figure \ref{Ms_UV1} except that the ``possible'' $1.1<z<1.9$ 
galaxies whose 90\% confidence ranges of photometric redshifts  
overlap the bin range are also plotted.
\label{z1test}}
\end{figure}
\begin{figure}
\epsscale{0.6}
\plotone{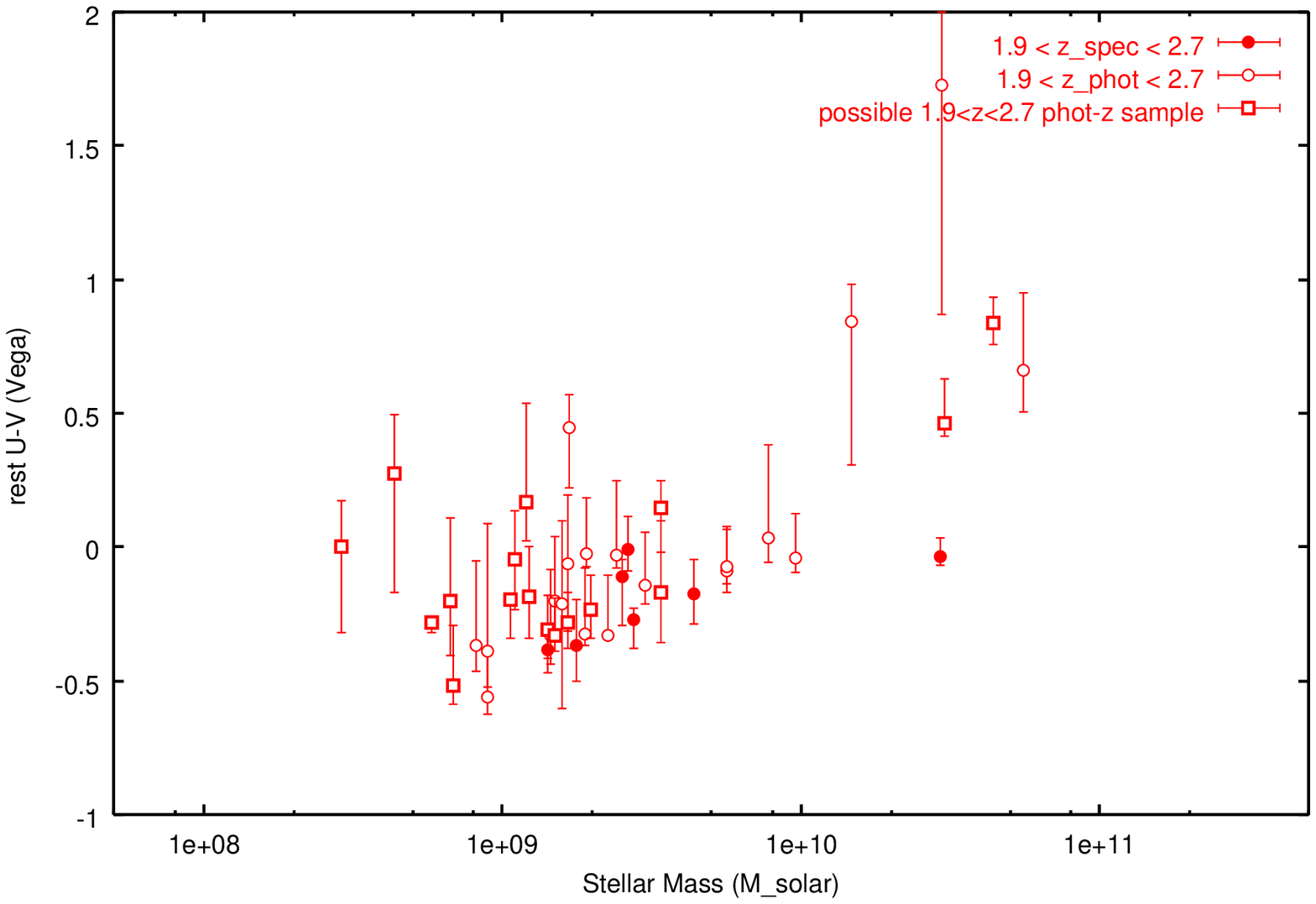}
\caption{same as Figure \ref{z1test}, but for $1.9<z<2.7$. 
\label{z2test}}
\end{figure}
Since our photometric redshift estimate seems to be relatively uncertain 
at $z \gtrsim 1.5$ (Figure \ref{testz}), we checked the effects of 
photometric redshift uncertainty on the membership of the 
 high-redshift bins. 
To do this, we picked up the objects without the spectroscopic 
redshift for which the 
 best-fit photometric redshifts are not included in the
considering redshift bin but the redshift ranges of their 90\%
confidence level in the SED fitting overlap the redshift bin
as ``possible sample''. 
Then we restricted their redshift within the considering redshift bin 
and evaluated the allowable ranges and the minimum $\chi^{2}$ values 
of the stellar mass and the rest $U-V$ color. 
The results for $1.1<z<1.9$ bin and $1.9<z<2.7$ bin 
are showed in Figure \ref{z1test} and \ref{z2test}. 
In these figures, ``possible sample'' does not affect the 
distribution of galaxies in the stellar mass vs rest $U-V$ plane
significantly.
Some objects whose best-fit photometric redshift is included 
 in the redshift bin could similarly escape from the bin, but such  
uncertainty about the photometric redshift does not seem to change the 
results obtained above.
\clearpage
\subsection{Morphology vs Stellar Mass}

In the previous subsection, we found that from the behavior of the
rest-frame $U-V$ color distribution, galaxies at $z\lesssim2$ can be
divided into two populations at around the stellar mass of $\sim
5\times10^{9}$M$_{\odot}$. In this section,
we examine the relationship between these 
populations divided by stellar mass and their morphology.
\begin{figure}
\epsscale{0.75}
\plotone{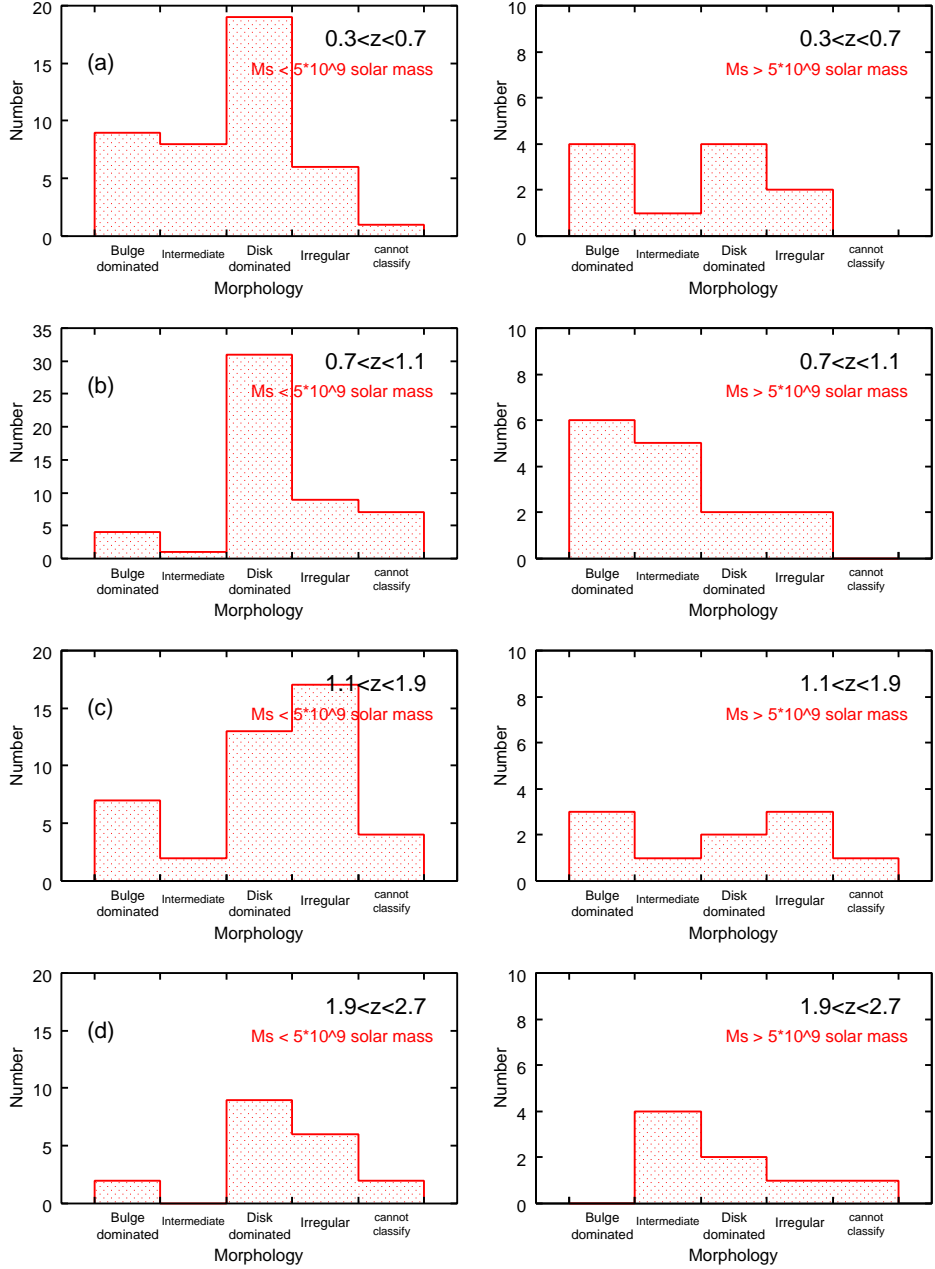}
\caption{Morphological distribution of our $K'$-selected galaxies
at $0.3<z<2.7$ in the HDF-N. left: for galaxies with
    with M$_{stellar}<5\times10^{9}$M$_{\odot}$. right: for galaxies
    with M$_{stellar}>5\times10^{9}$M$_{\odot}$.
     Each row corresponds to
each redshift bin. (a):$0.3<z<0.7$, (b):$0.7<z<1.1$, (c):$1.1<z<1.9$, 
(d):$1.9<z<2.7$.
\label{MORhist}}
\end{figure}

In Figure \ref{MORhist}, we show the morphological distribution of
the $K'$-selected galaxies in the HDF-N at each redshift bin with
M$_{stellar}<5\times10^{9}$M$_{\odot}$ and
M$_{stellar}>5\times10^{9}$M$_{\odot}$ separately.
From Figure \ref{MORhist}a and \ref{MORhist}b, 
it is seen that disk galaxies 
dominate the low-mass population at $z\lesssim1$. In the high-mass
sample, the various morphological types have similar fraction, and
the fraction of bulge-dominated galaxies seems to be higher than that
in the low-mass sample, although the small number of the high-mass
sample causes large uncertainty. At $1.1<z<1.9$, the number of irregular
galaxies increases and becomes similar with that of disk-dominated
galaxies in the low-mass sample, while the bulge-dominated galaxies
and irregular galaxies seem to have similar number in
the high-mass sample. In our morphological classification, 
since the galaxies classified into irregular category show the significantly
high asymmetry index relative to those of artificial galaxies with
similarly faint brightness, it is not the case that fainter galaxies
tends to be classified into the irregular category because of the noise
effect.
At $1.9<z<2.7$, the similar morphological
fraction with that at $1.1<z<1.9$ is seen.
\begin{figure}
\epsscale{0.7}
\plotone{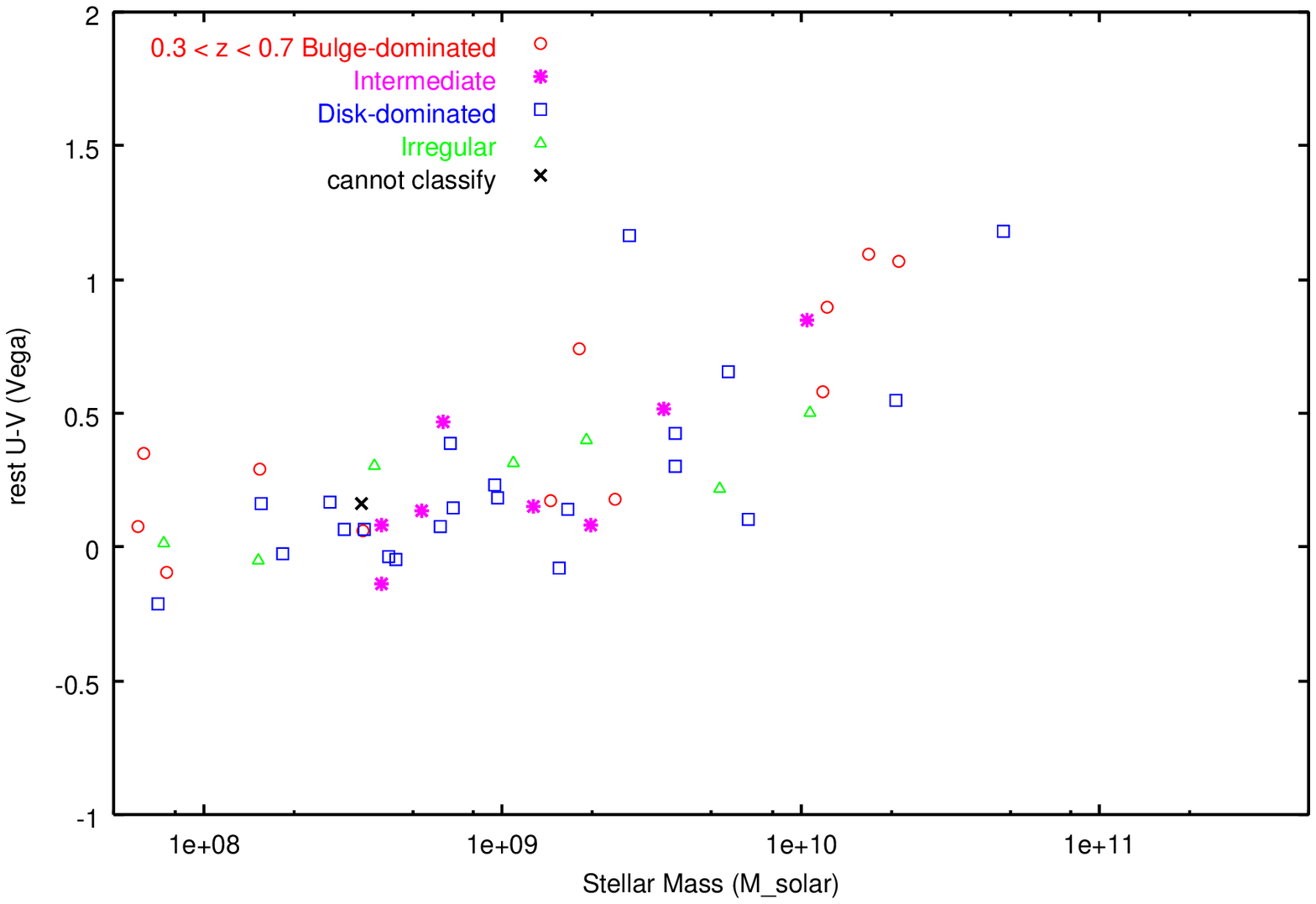}
\caption{$U-V$ distribution of galaxies with each morphology at
    $0.3<z<0.7$ as a function of stellar mass. Each symbol
    represents morphological type. Circles:bulge-dominated galaxies,
    Asterisks:intermediate, Squares:disk-dominated galaxies,
    Triangles:irregular galaxies, crosses:galaxies cannot be
    classified because of their faintness.
\label{UVMor03}}
\end{figure}
\begin{figure}
\epsscale{0.7}
\plotone{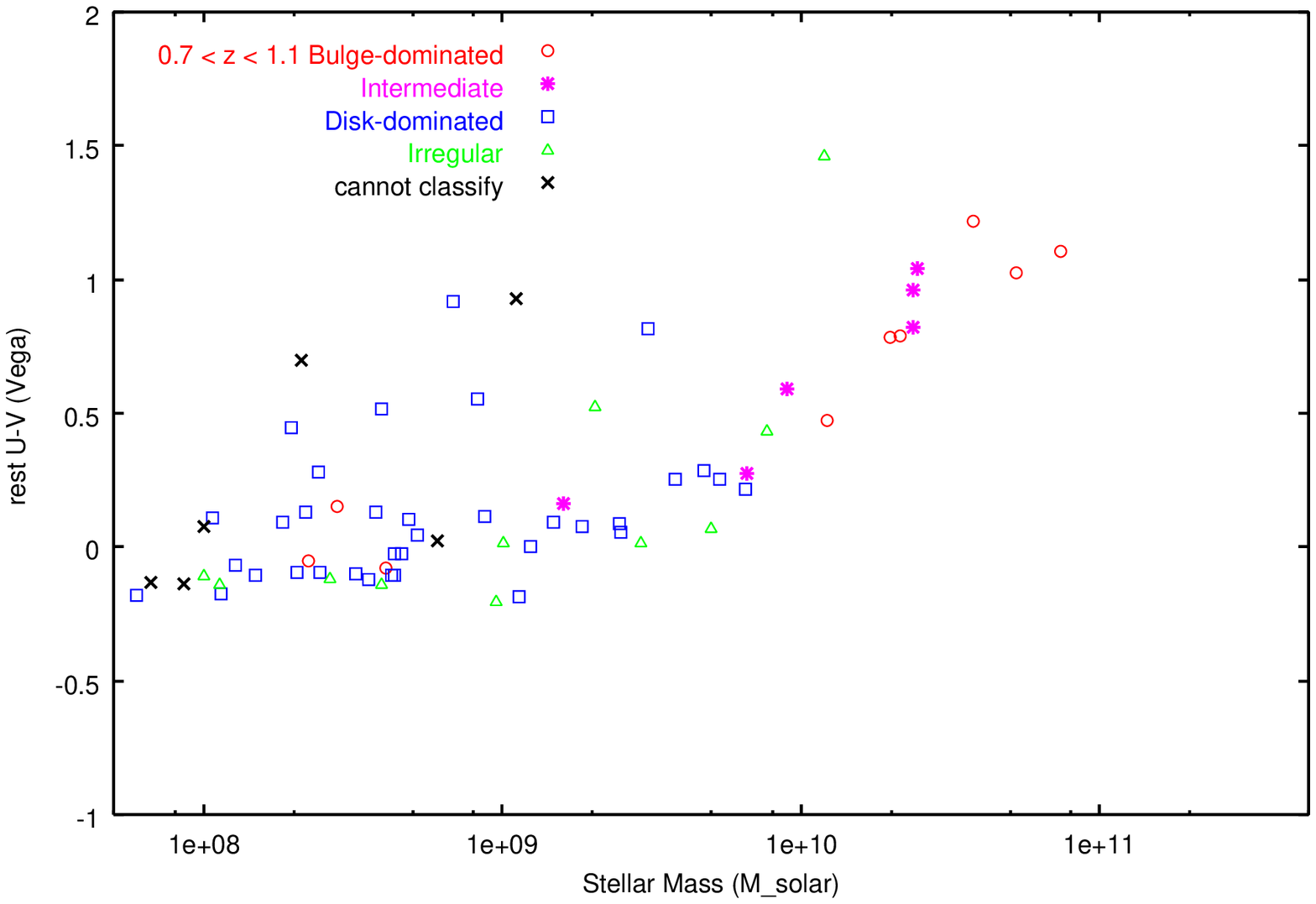}
\caption{Same as Figure \ref{UVMor03}, but for galaxies at $0.7<z<1.1$.
\label{UVMor07}}
\end{figure}
\begin{figure}
\epsscale{0.7}
\plotone{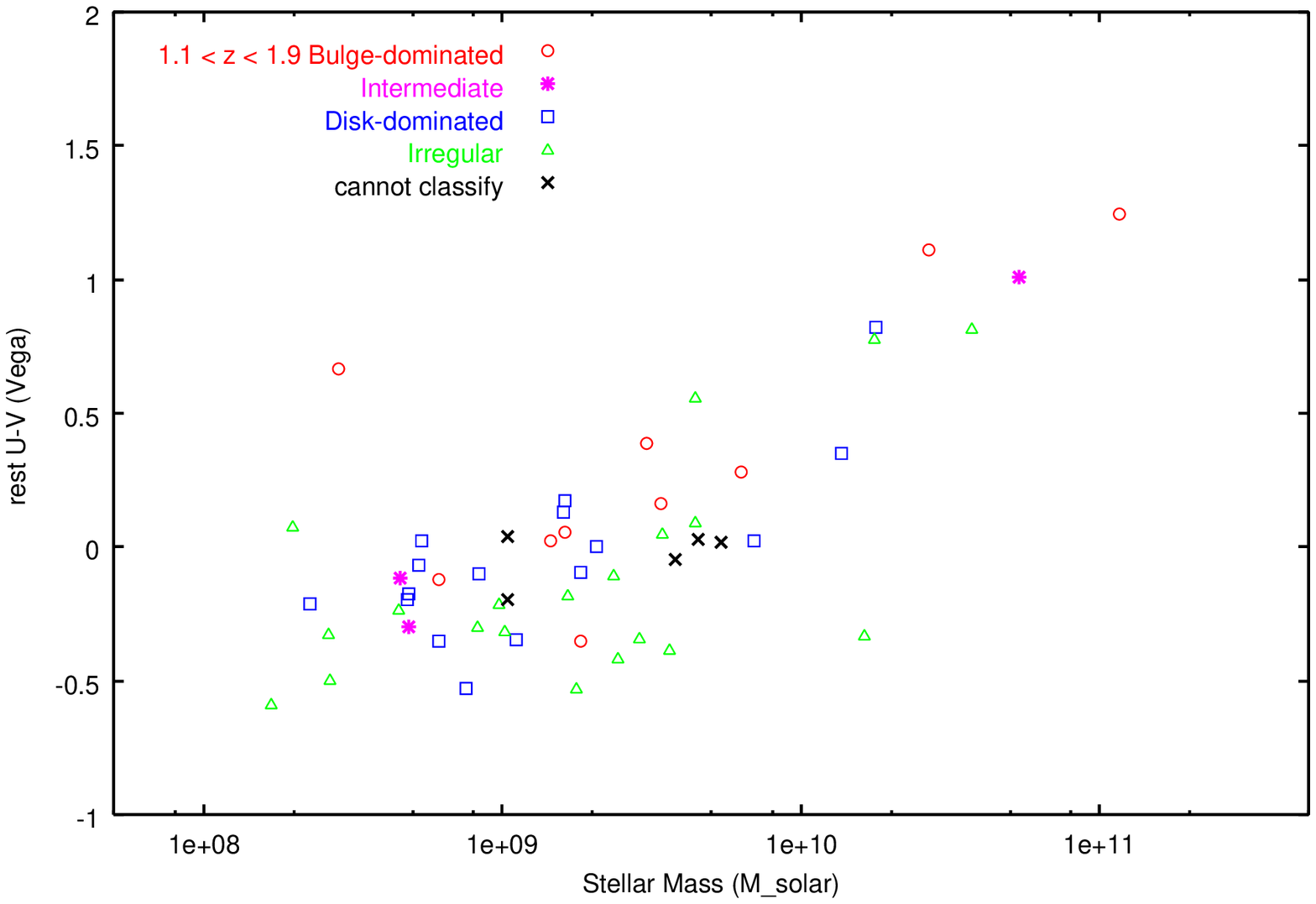}
\caption{Same as Figure \ref{UVMor03}, but for galaxies at $1.1<z<1.9$.
\label{UVMor1}}
\end{figure}
\begin{figure}
\epsscale{0.7}
\plotone{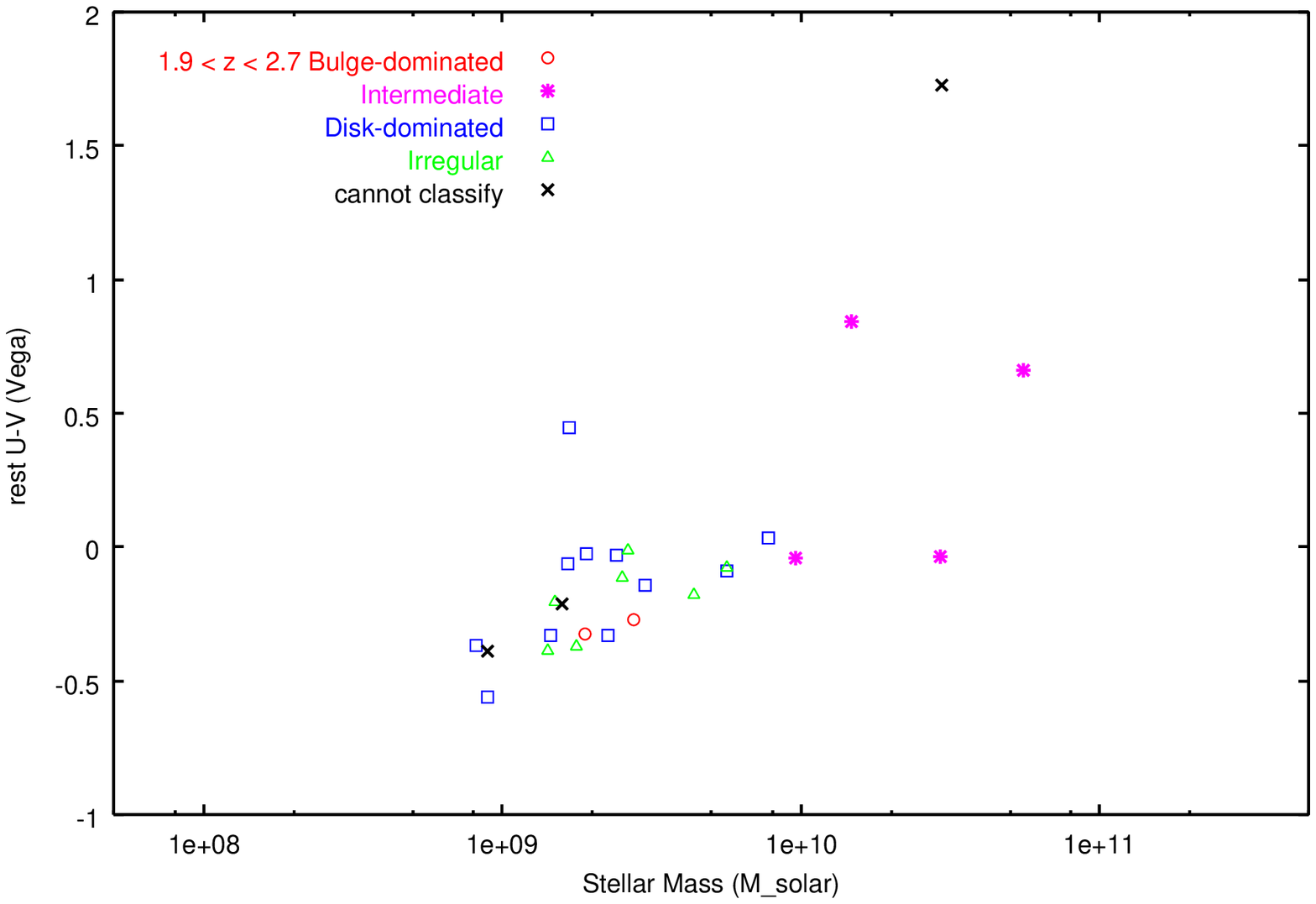}
\caption{Same as Figure \ref{UVMor03}, but for galaxies at $1.9<z<2.7$.
\label{UVMor2}}
\end{figure}

Figure \ref{UVMor03}-\ref{UVMor2} show the rest $U-V$ color 
distribution of galaxies 
with each morphological type as a function of stellar
mass. We also see here the trend that disk-dominated galaxies 
(and irregulars at $z\gtrsim1$) dominate
the low-mass population, and that in the high-mass population,
the fraction of galaxies with earlier-type morphology (bulge-dominated
and intermediate) is larger than that in the low-mass sample 
at $0.3\lesssim z\lesssim2.7$.  
Furthermore, in the stellar mass range of M$_{stellar}\gtrsim 5\times
10^{9}$M$_{\odot}$, there seems to be the trend that the more massive
galaxies have the earlier morphology.
\begin{figure}
\epsscale{0.78}
\plotone{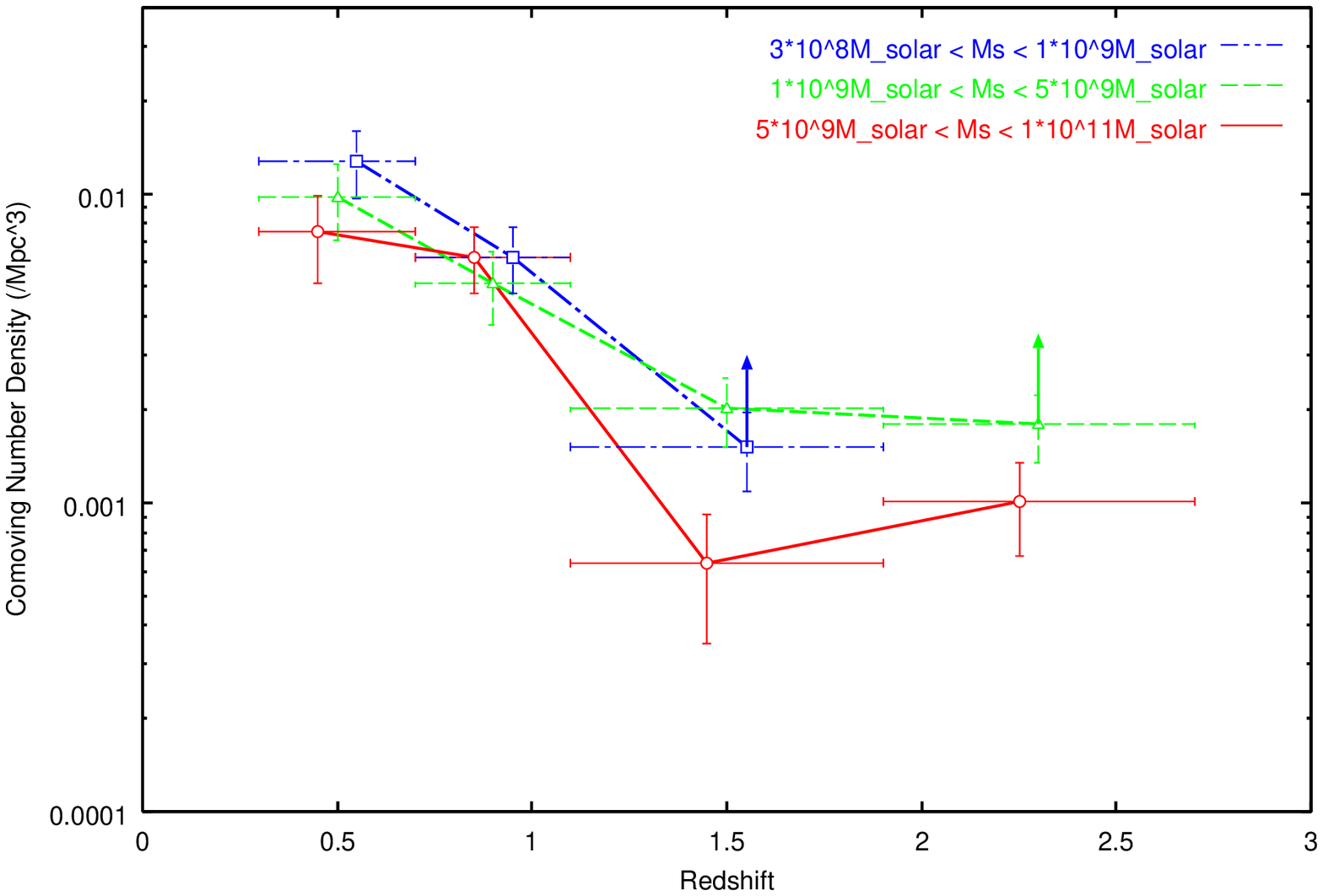}
\caption{Comoving number density of galaxies in each stellar mass range as a
    function of redshift. Each symbol represents the stellar mass
    range. Error bars are based on the square root of the observed
number. Arrows show that these bins are near the detection limit and 
should be regarded as lower limit considering the incompleteness. 
\label{ND_Ms}}
\end{figure}

In \citet{kaj01}, we investigated the morphological number
density evolution of galaxies with M$_{V}<-20$ at $z<2.0$ in the HDF-N, 
and found that the number density and morphological fraction of
bulge-dominated galaxies decreases at $z>1$.
\citet{sta04} also studied the properties of 
HST/NICMOS-selected early-type galaxies in
the HDF-N and found that the number density of morphologically
selected early-type galaxies decreases at $z\gtrsim1.4$.
On the other hand, 
in Figure \ref{MORhist}, each morphological fraction in the high and
the low-mass samples does not show significant change at $0.3\lesssim
z\lesssim2.7$, except that the fraction of irregular galaxies
increases at $z\gtrsim1$. In order to consider the consistency between
these results, 
in Figure \ref{ND_Ms}, we show the number density evolution of each
range of stellar mass. 
In the figure, it is seen that the number
density of the high-mass galaxies 
decreases rapidly at $z>1$ relative to
that of the low-mass sample in the HDF-N, although the uncertainty is
relatively large. 
It is noted that since we did not correct for the incompleteness in
this figure, the data of $1.1<z<1.9$ bin for the stellar mass range of
$3\times10^{8}$-$1\times10^{9}$M$_{\odot}$ and $1.9<z<2.7$ bin for
$1\times10^{9}$-$5\times10^{9}$M$_{\odot}$, which are near the
detection limit of our sample, should be considered as
lower limit value (arrows in the figure), 
and the difference of the number density decreases
at $z>1$ between for galaxies with M$_{stellar}>5\times
10^{9}$M$_{\odot}$ 
and with M$_{stellar}<5\times 10^{9}$M$_{\odot}$ could become larger
than that seen in Figure \ref{ND_Ms}.
Since the fraction of early-type
morphology is larger in the high-mass sample as mentioned above, 
in the overall mass range, 
the fraction of early-type is expected to decrease at $z>1$, and our
result about the morphological distribution as a function of 
stellar mass is consistent with the result in \citet{kaj01}.
\section{Discussion}
%

In the previous section, using the HST WFPC2/NICMOS archival data and very
deep Subaru/CISCO $K'$-band image of the Hubble Deep Field North,
we investigated the distribution of the rest
$U-V$ and morphology as a function of stellar mass back to
$z\sim3$. 
In the estimate of the stellar mass of each galaxy, even if each parameter
such as stellar age, star-formation timescale, or dust-extinction
 is not constrained so strongly, 
since the effects of these parameters tend to be
cancelled out as discussed above (see \citealp{pap01} for detailed
discussion),  we can evaluate the stellar mass with 
relatively high accuracy. 
Since the rest-frame $U-V$
color and rest-optical morphology are also measured from the observed 
$U$ to $K$-band images without extrapolation, our results obtained in
the previous section are based on the robust quantities. 
We here discuss some possible scenarios 
about galaxy formation and evolution 
inferred from the obtained results.
We consider the change of 
the mass dependence of the rest-frame color at the characteristic 
stellar mass first, and then discuss the evolution of the low-mass and the
high-mass population, respectively.  

\subsection{Characteristic Stellar Mass \label{chrms}}
First, we found that over $0.3<z<2$, 
the rest $U-V$ dependence on the stellar mass 
changed at M$_{stellar}\sim 5\times 10^{9}$M$_{\odot}$; in the 
high-mass region, the rest $U-V$ color of galaxies is strongly
correlated with the stellar mass and the trend that the more massive
galaxies have the redder $U-V$ color is seen. 
In the low-mass region, on the other hand, 
the $U-V$ color does not change so much along the stellar mass.  
Because the $U-V$ color is sensitive to the star formation histories, 
this indicates that the star formation activities of galaxies
are related with their stellar mass at each epoch.\\
In our analysis, the stellar mass at which this change of the mass 
dependence of the rest-frame color occurs does not change
significantly to $z\sim2$. Since most galaxies with
M$_{stellar}\lesssim 5\times10^{9}$M$_{\odot}$ have similarly blue $U-V$
color, which indicates active star formation,  
therefore the characteristic stellar mass   
for some mechanism which suppresses the star formation activities 
seems to be independent of redshift at $0.3\lesssim z\lesssim2$.

For local galaxies, 
\citet{kau03} reported the similar 
dependence of star formation histories on the stellar mass 
based on the Sloan Digital Sky Survey data. In their analysis, 
galaxies divide into two populations at M$_{stellar
}\sim3\times10^{10}$M$_{\odot}$, which is higher value than 
that we found for galaxies in the HDF-N. 
As seen from their Figure 1 (D$_{n}$4000 or H$\delta$
vs stellar mass), however, while their low-mass population spreads
over the range of 1$\times10^{8}$-5$\times10^{9}$M$_{\odot}$, the
high-mass population has the peak at a stellar mass larger than
$\sim5\times10^{11}$M$_{\odot}$, where we detected few galaxies in
the HDF-N. Therefore the high-mass population in our analysis
corresponds to the transition zone in Kaffumann et al.'s analysis,
and the trend that  
the correlation between star formation activity and the stellar mass 
changes at $\sim5\times10^{9}$M$_{\odot}$
is also seen for local galaxies in the SDSS data.

\begin{figure}
\epsscale{0.85}
\plotone{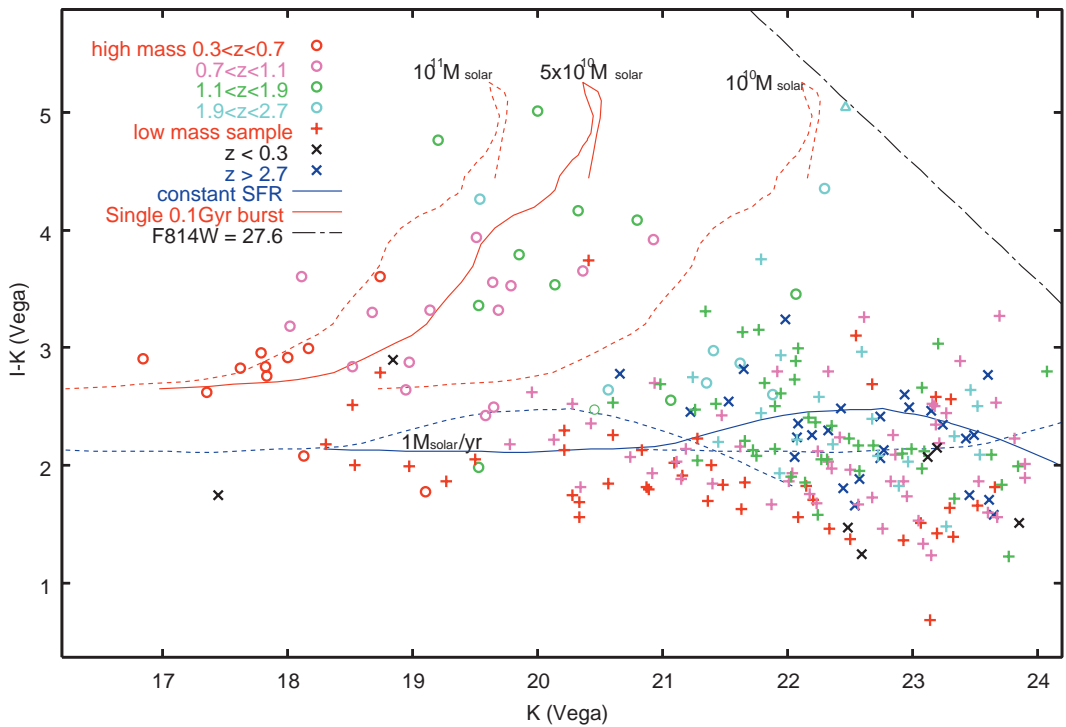}
\caption{Observed $I_{814}-K$ vs $K$ color-magnitude diagram for our
    $K'$-selected sample in the HDF-N.
    Circles represent 
    galaxies with M$_{stellar}>5\times10^{9}$M$_{\odot}$,
    while crosses show galaxies with 
    M$_{stellar}<5\times10^{9}$M$_{\odot}$.
    Color of each symbol represents its redshift.
    Objects without the redshift range of our analysis (i.e. objects
    with $z<0.3$
    or $z>2.7$) are showed as X symbols (black for $z<0.3$, blue for $z>2.7$).
    For comparison, the 0.1 Gyr single burst models(red lines)
 and constant star
    formation rate models (blue lines)
are showed as sequences of redshifts ($0.3<z<2.7$). 
    These models assume solar metallicity and no dust extinction, and
    formation redshift of 3.5. For single burst model, the models with
    a stellar mass of 1$\times10^{11}$M$_{\odot}$(dashed line),
    5$\times10^{10}$M$_{\odot}$(solid line),
    1$\times10^{10}$M$_{\odot}$(dashed line) are showed.
    For constant star formation models, the models with a star formation
    rate of 0.1M$_{\odot}$ yr$^{-1}$ (dashed), 1M$_{\odot}$ yr$^{-1}$ (solid),
    10M$_{\odot}$ yr$^{-1}$ (dashed) are showed.
\label{IK_K}}
\end{figure}
At $1.9<z<2.7$, the number of galaxies with
M$_{stellar}> 1\times10^{10}$M$_{\odot}$ in our sample becomes very
small. Furthermore,  
our detection limit corresponds to $\sim1\times10^{9}$M$_{\odot}$ at
this redshift range.
Therefore we cannot confirm whether the change of the dependence of $U-V$
color on the stellar mass occur or not, although we found that 
galaxies with higher stellar mass tend to have redder rest
$U-V$ color in the mass range of
$\sim1\times10^{9}$--1$\times10^{10}$M$_{\odot}$.
\citet{fon03} found that 
in addition to the star forming galaxies, there are a
few red galaxies at $z\gtrsim2$ in the HDF-S, and that as the stellar mass
increases, the fraction of older objects seems to increase, although
the biases against old/passive objects exist at low-mass region.
These results suggest that the mechanism which suppresses the star
formation in high-mass galaxies may work at these 
earlier epochs.

The difference of color distribution between the low and high-mass
samples is also seen in the observed color--magnitude diagram, which
is {\it not} affected by the uncertainty of redshift determination.
Figure \ref{IK_K} shows the observed $I_{814}-K$ vs $K$ diagram for our 
$K'$-selected sample in the HDF-N. The symbols represent their stellar 
masses: circles represent galaxies with M$_{stellar}>
5\times10^{9}$M$_{\odot}$, and crosses show galaxies with M$_{stellar}<
5\times10^{9}$M$_{\odot}$. Objects without the redshift range of our
analysis are showed as X symbols. 
Color of each symbol shows its redshift. 
In this figure, for comparison, the 0.1Gyr single burst (thereafter 
passively evolving) models with a stellar mass of 
1$\times10^{10}$M$_{\odot}$, 5$\times10^{10}$M$_{\odot}$,
1$\times10^{11}$M$_{\odot}$ and the constant star formation models 
with a star formation rate of 0.1M$_{\odot}$ yr$^{-1}$, 1M$_{\odot}$ yr$^{-1}$,
10M$_{\odot}$ yr$^{-1}$ are showed as sequences of redshift between 0.3 and
2.7. The distribution of galaxies in the HDF-N in
$I_{814}-K$ vs $K$ diagram can be divided into two groups: the one
group is spread near the passive evolution models, the other has
fainter $K$-magnitude and bluer $I_{814}-K$ color, and is spread
around the constant star formation models. The gap between two groups 
seems to lie roughly along the single burst model with a stellar mass
of 10$^{10}$M$_{\odot}$. This reflects the result discussed above 
that the change of 
the rest $U-V$ color distribution at a stellar mass of
$\sim5\times10^{9}$M$_{\odot}$ is seen at $0.3\lesssim z \lesssim2$.
 The similar gap in the $I_{814}-K$ vs $K$ diagram is also seen in the
Hawaii Deep Survey \citep{cow95}, or 53W002 field \citep{yam01}.
\begin{figure}
\epsscale{0.65}
\plotone{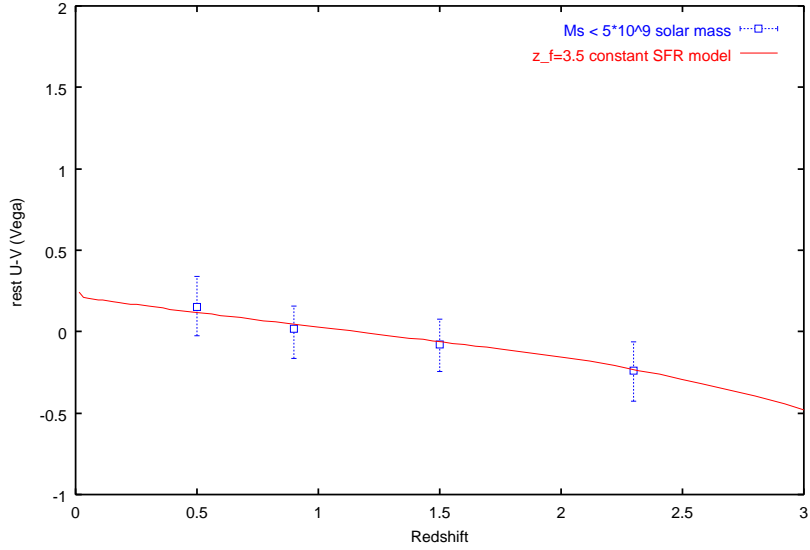}
\caption{Rest-frame $U-V$ color of constant star
    formation rate model with formation redshift of 
3.5 as a function of redshift. Squares represent the average $U-V$
    color of galaxies with M$_{stellar}<5\times10^{9}$M$_{\odot}$ at
    each redshift bin. Errorbars show the root mean square 
 around the average value.
\label{Const}}
\end{figure}

\subsection{The Low-mass Population}
Second, we found that while the $U-V$ color of galaxies with 
M$_{stellar}<5\times10^{9}$M$_{\odot}$ is only weakly correlated with 
their stellar mass, the average of their $U-V$ color becomes bluer 
with redshift from $U-V\sim0.2$ at $z\sim0.5$ to $U-V\sim-0.2$ at
$z\sim2$. Their bluer color of galaxies at higher redshifts
indicates that their average stellar age is younger. 
On the other hand, 
the $U-V$ color of the low-mass sample at $0.3<z<0.7$ ($U-V\sim0.2$)  
 indicates that active star formation still occurs in these galaxies. 
Since the most of the galaxies with 
M$_{stellar}<5\times10^{9}$M$_{\odot}$ have such blue $U-V$ colors at 
any redshifts between $z\sim0.3$ and $z\sim2.7$, these galaxies seem to have 
relatively long characteristic timescale of star formation, although 
in our SED fitting in section \ref{smass}, the star formation timescale
$\tau$ of each galaxy at each redshift cannot be constrained strongly. 

If we consider constant SFR model,   
the color of $U-V\sim -0.2$ at $z\sim2$ indicates that their stellar age 
is $\lesssim$1Gyr old, and their 
formation redshift does not seem to be much higher than the observed epoch. 
For example, Figure \ref{Const} shows the rest $U-V$ color of the constant star
formation rate model with formation redshift of 3.5 calculated with GALAXEV
code, as a function of redshift. 
$H_0$ = 70 km s$^{-1}$ Mpc$^{-1}$,
$\Omega_{\rm 0}=0.3$, $\Omega_{\Lambda}=0.7$ cosmology and solar
metallicity are assumed. Such a simple model can
reproduce the $U-V$ color distribution of the low-mass sample at each
redshift relatively well.

\begin{figure}
\epsscale{0.65}
\plotone{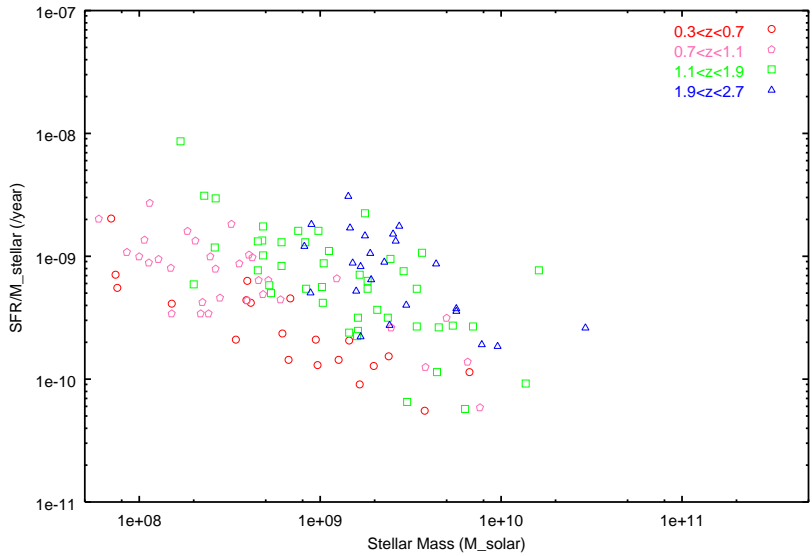}
\caption{Estimated star formation rate 
relative to stellar mass of galaxies with $U-V<0.5$ as
    a function of stellar mass. Symbols represent corresponding
redshift bins.
\label{SSFR_Ms1}}
\end{figure}

If we assume the continuous star formation for these low-mass
galaxies as discussed above, to  
what extent these star formation activities can grow
their stellar mass ? 
Since in our SED fitting procedure, where star formation
time-scale $\tau$, age, and dust-extinction are free parameter, the  
star formation rate or star formation time-scale of each galaxy can
be only very weakly constrained, here we assume simply the constant star
formation rate ($\tau=\infty$) and no-extinction in order to do the rough
estimation of the star formation rate of each galaxy.
We derived the 
star formation rate of these galaxies from the observed photometry
which corresponds to the rest-frame 1500-2000 \AA, comparing
 the GALAXEV model with 
constant SFR, solar metallicity, Chabrier et al.'s IMF. 
Under these assumptions, the absolute magnitude at $\sim$1500-2000\AA\   
 is insensitive to the age at larger than 0.5 Gyr old, and seems to
be the good indicator of star formation rate. The results are showed
as the star formation rate relative to stellar mass, namely
specific star formation rate, in Figure \ref{SSFR_Ms1}. 
From the figure, it is seen that at the same 
stellar mass, the star formation rate become gradually higher as
redshift increases. It is noted that at
M$_{stellar}\lesssim 10^{9}$M$_{\odot}$, galaxies at $z\gtrsim 2$ are
not selected into our sample because of the detection limit of our
observation. 
At $z\gtrsim1$, the galaxies
with a stellar mass of $\sim 1\times 10^{9}$M$_{\odot}$ form stars at
the rate of 0.5--1M$_{\odot}$ yr$^{-1}$, 
which corresponds that the stellar
mass of these galaxies would become about 1.5--2.0 times larger 
after 1Gyr. 
At $0.7<z<1.1$, the star formation rate of those with 
M$_{stellar}\sim 1\times 10^{9}$M$_{\odot}$
becomes slightly smaller, and
the stellar mass of these galaxies become no more than about 1.2--1.5 times
larger per 1Gyr. 
The smaller stellar mass galaxies seem to have higher
specific star formation rate. Those galaxies with M$_{stellar}\sim
1\times 10^{8}$M$_{\odot}$ will have about twice stellar
mass at 1 Gyr after if star formation rate would continue to be
constant. 
The growth rate of stellar mass is still smaller at $0.3<z<0.7$. 

On the other hand, the specific star formation 
rate represents the inverse of the age of the objects  
under the assumption of the constant SFR. We can compare these ages 
estimated from the star formation rate and the stellar mass with 
the formation epoch inferred from the rest-frame color evolution.
For example, the constant SFR model with formation redshift of 3.5, 
which can explain the evolution of the average $U-V$ color well in Figure 
\ref{Const}, becomes 1Gyr old at $z\sim2.3$, 2.4Gyr at 
$z\sim1.5$, 4.4Gyr at $z\sim0.9$, and 6.6Gyr at $z\sim0.5$, respectively.
The corresponding specific star formation rate becomes 
from $1\times10^{-9}$ yr$^{-1}$ at $z\sim2.3$, 
to $1.5\times10^{-10}$ yr$^{-1}$ at $z\sim0.5$.
In Figure \ref{SSFR_Ms1}, however, 
 the specific star formation rates estimated from the rest-UV 
luminosity   
are wide-spread in each redshift bin, and the star formation rates  
 of some galaxies seem to be inconsistent with the formation epoch 
inferred from the rest $U-V$ color evolution. 
For example, the galaxies with
M$_{stellar}\sim1\times10^{8}$M$_{\odot}$ at $0.7<z<1.1$ have 
the specific star formation rate of $\sim 1\times10^{-9}$ yr$^{-1}$, and 
their formation epoch is expected to 
be about 1Gyr before the observed time, which
corresponds to $z\sim$1-1.5. 
These discrepancies may be explained by the following factors 
which could affect our estimation of 
the star formation rate and the mass growth rate.\\
So far, several studies of Lyman break galaxies at $z\gtrsim2$ suggested
that they are shrouded in some amount of dust, typically
E(B-V)$\sim0.15$ with relatively large scatter 
(e.g., \citealp{ste99,pap01,sha01}).  
In fact, the best-fit E(B-V) of these low-mass blue galaxies in our sample
has similar distribution with large scatter, although 
we can only weakly constrain the amount of the extinction of these galaxies. 
If we consider such an amount of dust extinction for these low-mass
galaxies, the estimated star
formation rates could become several times to an order of magnitude larger.
\\
Several studies also suggested that the star
formation in Lyman break galaxies is recurrent with relatively short
timescale \citep{pap01,sha01}. \citet{gla99} also pointed out that the 
H$\alpha$ and UV luminosity of  
star forming galaxies at $z\sim1$ indicates their star formation 
occurs in episodic bursts. 
Since most low-mass galaxies in our sample have rather blue rest $U-V$ color 
at any redshifts between $z\sim0.3$ and $z\sim2.7$ as mentioned above, 
the duty cycle of star formation should have relatively short timescale
(e.g., $\lesssim0.5$-1Gyr) if we assume those recurrent star formation
activities. In this case, the expected stellar mass growth of these
galaxies depends on the period of the duty cycle. 
Therefore, the formation epoch calculated from the stellar mass and the star
formation rate at the observed time under the assumption of the
constant SFR may not necessarily corresponds to that inferred from the
color evolution.

With regard to morphological distribution, while disk-dominated
galaxies dominate this low-mass population at $z\lesssim1$, the fraction of
irregular galaxies increases at $1.1<z<1.9$. The $U-V$ color
distribution of irregular galaxies at $0.7<z<1.9$ is similar to, 
 or slightly bluer than that of disk-dominated galaxies.  
If we assume that most of these irregular galaxies are merger origin
(or interactions), these processes do not seem to change their colors.
Namely, such processes may not ceases the star formation activities in
these low-mass galaxies.
\subsection{The High-mass Population}
Third, we found that the correlation between the stellar mass and the
rest $U-V$ color of galaxies with
M$_{stellar}\gtrsim 5\times10^{9}$M$_{\odot}$ does not show significant
evolution between $0.3\lesssim z \lesssim2$.
These high-mass galaxies have redder color, and earlier type
morphology than the low-mass sample. 
For galaxies at $0.2<z<1.0$, \citet{bri00} estimated the stellar
mass of galaxies with spectroscopic redshift, using the Optical and NIR
photometries from the Canada France Redshift Survey data, and
investigated the relation between the stellar mass and the specific
star formation rate. 
At the range of stellar mass of
$9.5\lesssim$logM$_{stellar}\lesssim12.0$, which corresponds to their
sample selection of $17.5<I<22.5$, they found that the more
massive galaxies tend to have the smaller specific star formation rate
and that these correlations  
evolve only slightly to more active star-forming between $z\sim0.2$
and $z\sim1.0$ 
(their Figure 3). Since the specific star formation rate
is considered to be related with the  
rest $U-V$ color estimated in our analysis, 
 at $0.3\lesssim
z\lesssim1.0$ and the range of the stellar mass of
$5\times10^{9}$--$1\times10^{11}$M$_{\odot}$, our result is
qualitatively consistent with that of \citet{bri00}. 
While Brinchmann \& Ellis's result indicates that this correlation
continues to the stellar mass of $\sim1\times10^{12}$M$_{\odot}$ at
least at $z\lesssim1$, our result suggests that this relation holds to
$z\sim2$.

\begin{figure}
\epsscale{0.6}
\plotone{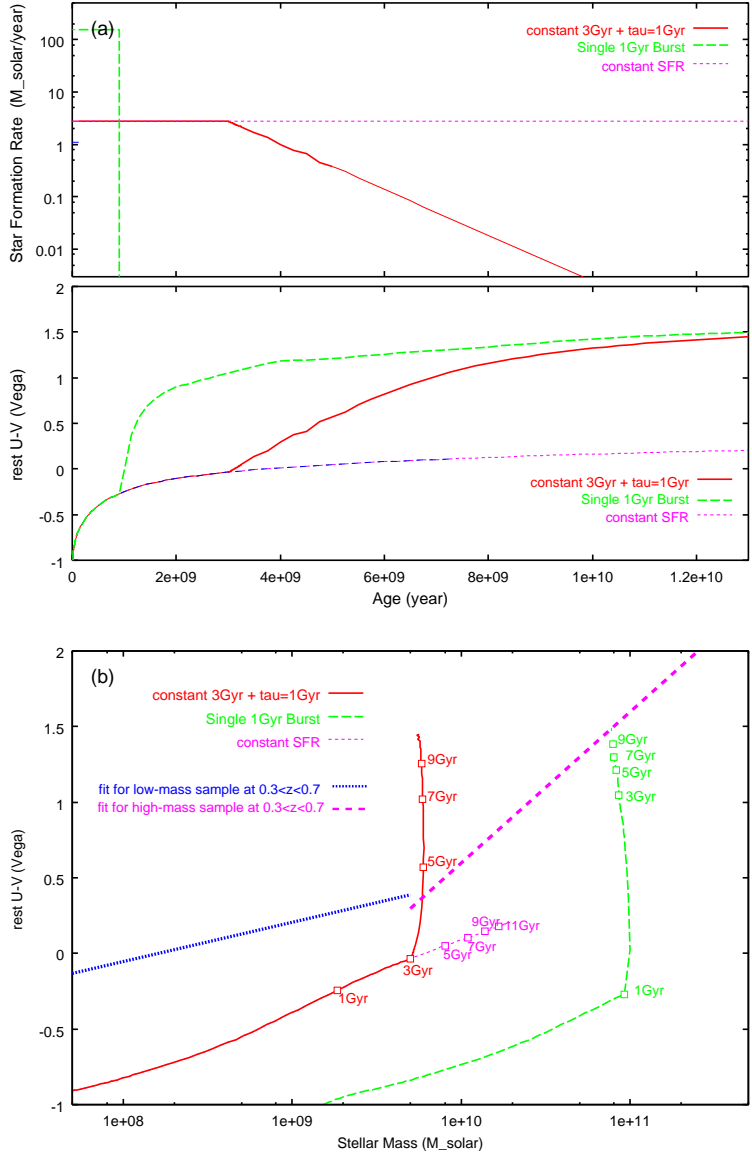}
\caption{(a): Top: star formation rate vs age for the models with various star
    formation history. Constant 3Gyr SFR $+ \tau=$1Gyr model with
    5$\times 10^{9}$M$_{\odot}$ at 3Gyr old (solid), single 1Gyr
    burst with 10$^{11}$M$_{\odot}$ at 1Gyr old (dash), constant SFR with
    5$\times 10^{9}$M$_{\odot}$ at 3Gyr old (short-dashed). 
Bottom: rest $U-V$ color
    vs age for the same models calculated from the GALAXEV library.
    (b): rest $U-V$ color vs stellar mass for the models in (a). 
    Each line corresponds to each star formation history of (a). Each
square represents age. Bold lines show the fitting result for the
observed galaxies at $0.3<z<0.7$. 
\label{Ms_UVmodel}}
\end{figure}
What kind of the galaxy evolution scenario can explain this
trend ?
If we consider the pure passive
evolution for these galaxies, the $U-V$ color is expected to be bluer 
at higher redshift with nearly constant stellar mass.
This seems to disagree with our result. 
For example, in Figure \ref{Ms_UVmodel}, 
we show the track of such a  model 
in the rest $U-V$ color vs stellar mass plane, using 
the single 1Gyr burst model with 10$^{11}$M$_{\odot}$ at
1Gyr old (dashed line). As seen from the model, 
if observed high-mass galaxies follow passive evolution, 
the correlation between the $U-V$ color and the stellar mass 
shifts blue-ward with redshift especially at low mass side  
unless the degree of dust extinction 
is regulated with the stellar mass and the stellar age. 
In order to make the relation between the stellar mass and the
rest $U-V$ color nearly constant over relatively long timescale, galaxies
will have to increase their stellar mass as their stars become
older.  
Furthermore, since the comoving volumes of $1.1<z<1.9$ bin and
$1.9<z<2.7$ bin are more than three times larger than that of
$0.7<z<1.1$ bin, the number density of galaxies with
M$_{stellar}\gtrsim1\times10^{10}$M$_{\odot}$ seems to decrease
between $z\sim1$ and $z\sim3$, although small number statistics
prevents the conclusive result. For relatively bright (probably massive)
early-type galaxies, several
previous studies also found that the number density of these galaxies 
seems to decrease at $z\gtrsim1.4$
(\citealp{zep97,fra98,bar99,men99,tre99,sta04}, see also
\citealp{dad00,mac01}).\\  
The merging without active star formation may
be possible solution to increase the stellar mass of these galaxies. 
In this scenario, the observed relation corresponds that 
the stellar mass increases through mergers from 
$\sim5\times10^{9}$M$_{\odot}$ to $\sim5\times10^{10}$M$_{\odot}$
 during the time when the
$U-V$ color evolves from $\sim0.4$ to $\sim1.2$.
For example, in Figure \ref{Ms_UVmodel}, we assume the model where the star
formation rate is constant by the time the stellar mass reaches 
5$\times 10^{9}$M$_{\odot}$, and then star formation rate decreases
exponentially with characteristic time scale $\tau=$1Gyr (solid line). 
We consider the case that 
the period for which star formation rate is constant is 3Gyr.
Star formation rate is adjusted such that the stellar mass 
becomes  5$\times 10^{9}$M$_{\odot}$ at 3Gyr old,
 (Figure \ref{Ms_UVmodel}(a)). 
In the model, after 3Gyr old, the star formation
rate decreases and the stellar mass does not grow. Comparing this 
(no merger) model with the observed correlation between the rest $U-V$ and
stellar mass, we found that in order to form the observed correlation, 
galaxies have to increase their stellar mass after star formation
activities decrease at the rate of 2--3 times per about 2Gyr 
through the processes such as mergers 
for the assumed star formation history.

\section{Summary}
Using the HST WFPC2/NICMOS archival data and very deep 
Subaru/CISCO $K'$-band image 
of the Hubble Deep Field North, we estimated the stellar mass,
the rest $U-V$ color, the morphology at rest-frame optical band of
$K'$-selected galaxies, and investigated the distribution of the rest
$U-V$ and morphology as a function of stellar mass back to
$z\sim3$. Following results were obtained.

\begin{itemize}
\item In the rest $U-V$ color vs stellar mass diagram, galaxies can
 be divided into two populations at M$_{stellar} \sim 
5\times 10^{9}$M$_{\odot}$. Galaxies with M$_{stellar}\lesssim
5\times10^{9}$M$_{\odot}$ have relatively blue rest $U-V$ color and
their color does not correlate strongly with stellar mass, while
at M$_{stellar} \gtrsim 5\times10^{9}$M$_{\odot}$, there is the trend
that galaxies with higher stellar mass have redder $U-V$ color. This 
feature seems to hold back to $z\sim2$.

\item We could not find the significant evolution of the characteristic 
stellar mass
at which the change of the mass dependence of the rest $U-V$ color
occurs, at $0.3\lesssim z \lesssim 2$.

\item The $U-V$ color of the low-mass sample become bluer
gradually with redshift, from $U-V\sim0.2$ at $z\sim0.5$, to
$U-V\sim-0.2$ at $z\sim2$. Such a color evolution is roughly
consistent with that expected from the constant SFR model with a
formation redshift of $z_{f}\sim3.5$.

\item On the other hand, the strong correlation between
the rest $U-V$ color
and the stellar mass seen in the high-mass sample does
not show significant evolution at $0.3\lesssim z \lesssim2$, although
the number density of high-mass galaxies may decrease at
$z\gtrsim1$. 

\item At $z\lesssim1$, disk galaxies dominate the low-mass
population, while the fraction of early-type morphology is larger 
in the high-mass population. Although the fraction of irregular
galaxies increases, the same trend as $z\lesssim1$ is seen at $z\gtrsim1$.

\item At $z>2$, although we can only sample the galaxies with
M$_{stellar}\gtrsim 1\times10^{9}$M$_{\odot}$, it is seen that
galaxies with higher stellar mass tend to have redder rest $U-V$ color
over the mass range of $1\times10^{9}$--$1\times10^{10}$M$_{\odot}$.
\end{itemize}

Finally, these results about galaxy evolution are based on the HDF-N
data, which is at most 4 arcmin$^{2}$. The field-to-field variance 
for such a small volume analysis is inferred to be relatively large.
Although the $I-K$ vs $K$
color-magnitude diagram of several other fields suggests that 
the close connection between the star formation activities and the
stellar mass found in our analysis could be applied to other general
fields as mentioned in section \ref{chrms}, the 
investigation/confirmation by the larger-volume survey is clearly
important, especially to reveal the evolution of high-mass galaxies, 
which are relatively rare and maybe strongly clustering.

\acknowledgments
We would like to thank Ichi Tanaka for valuable discussion.
This paper is based on data collected using the Subaru telescope,
which is operated by the National Astronomical Observatory of Japan.
This work is based in part on observations with the NASA/ESA Hubble
Space Telescope, obtained from the data archive at the Space Telescope
Science Institute, U.S.A., which is operated by AURA, Inc. under NASA
contract NAS5-26555. 
This work is partially supported by the grants-in-aid for scientific
research of the Ministry of Education, Culture, Sports, Science, and
Technology (14540234).
The Image Reduction and Analysis Facility (IRAF)
used in this paper is distributed by National Optical Astronomy
Observatories, U.S.A., operated by the Association of Universities for
Research in Astronomy, Inc., under contact to the U.S.A. National
Science Foundation.

\clearpage

\begin{table}
\begin{center}
\caption{Slope value of the linear fit in the rest $U-V$ color vs
$\log{M_{stellar}}$ plane \label{slope}}

\begin{tabular}{ccc}
\tableline\tableline
Redshift & M$_{stellar}<5\times10^{9}$M$_{\odot}$ & M$_{stellar}>5\times10^{9}$M$_{\odot}$\\ 
\tableline
0.3-0.7 &0.260$\pm$0.096 &1.002$\pm$0.269\\
0.7-1.1 &0.152$\pm$0.095 &0.801$\pm$0.196\\
1.1-1.9 &0.049$\pm$0.119 &0.980$\pm$0.279\\
1.9-2.7 &0.450$\pm$0.546 &0.425$\pm$0.437\\
\tableline
\end{tabular}

%
\end{center}
\end{table}
\begin{table}
\begin{center}
\caption{Offsets of the average 
rest $U-V$ color from the value at $0.3<z<0.7$ \label{cept}}

\begin{tabular}{ccc}
\tableline\tableline
Redshift & M$_{stellar}<5\times10^{9}$M$_{\odot}$ & M$_{stellar}>5\times10^{9}$M$_{\odot}$\tablenotemark{a}\\ 
\tableline
0.7-1.1 &-0.134$\pm$0.223 &-0.115$\pm$0.274\\
1.1-1.9 &-0.234$\pm$0.230 &-0.224$\pm$0.362\\
1.9-2.7 &-0.394$\pm$0.254 &\\
\tableline
\end{tabular}

%
\tablenotetext{a}{Offsets of the 
intercept estimated from the linear fitting where the slope of
$0.3<z<0.7$ bin is assumed(see text).}

\end{center}
\end{table}

\end{document}